\documentclass[12pt]{amsart}

\usepackage{tabularx}
\usepackage{amsmath}
\usepackage{amssymb}
\usepackage{amsfonts}
\usepackage{boxedminipage}
\usepackage{appendix}
\usepackage{algorithm2e}
\RestyleAlgo{boxruled}
\usepackage{mathtools}
\usepackage{physics}
\usepackage{subfigure}
\usepackage{cite}

\usepackage{graphicx}
\usepackage[export]{adjustbox}

\usepackage{flexisym}	 
\usepackage{breqn}       

\usepackage{amsmath,amssymb,amsthm}
\usepackage{mathtools}   
\usepackage{physics}

\usepackage[utf8]{inputenc} 

\usepackage{csquotes}     
\MakeOuterQuote{"}       

\usepackage{fullpage}

\usepackage[colorinlistoftodos]{todonotes}   

\usepackage{hyperref} 
 \hypersetup{%
 	colorlinks=true,       
 	linkcolor=blue,          
 	citecolor=blue,        
 	filecolor=magenta,      
 	urlcolor=blue         
 }
\usepackage{amsaddr}
\DeclarePairedDelimiterX{\innerp}[2]{\langle}{\rangle}{#1,#2}
\renewcommand{\ip}[2]{\innerp*{#1}{#2}}

\newcommand{\vw}{\vb{w}}

\newcommand{\Matlab}{\textsc{Matlab}}

\newcommand{\Md}{M_{\rm d}}
\newcommand{\psid}{\psi_{\rm d}}

\newcommand{\CC}{{\mathbb C}}
\newcommand{\RR}{{\mathbb R}}

\numberwithin{equation}{section}
\numberwithin{figure}{section}

\graphicspath{{./Images/}}

\makeatletter
\renewcommand{\email}[2][]{%
  \ifx\emails\@empty\relax\else{\g@addto@macro\emails{,\space}}\fi%
  \@ifnotempty{#1}{\g@addto@macro\emails{\textrm{(#1)}\space}}%
  \g@addto@macro\emails{#2}%
}
\makeatother

\title{Efficient Manipulation of Bose-Einstein Condensates in a Double-Well Potential}

\author{$^\ast$J. Adriazola$^{(\MakeLowercase{a})}$}
\thanks{$^\ast$Corresponding author (Currently at the Universtiy of California, Santa Barbara).}

\author{R. H. Goodman$^{(\MakeLowercase{b})}$}
\address[a,b]{Department of Mathematical Sciences, New Jersey Institute of Technology, University Heights, Newark, NJ, USA, 07102}
\email[a]{jadriazola@ucsb.edu}
\email[b]{goodman@njit.edu}

\author[c]{P. G. Kevrekidis$^{(\MakeLowercase{c})}$}
\address[c]{Department of Mathematics and Statistics, University of Massachusetts Amherst, Amherst, MA, USA, 01003-4515}
\email[c]{kevrekid@umass.edu}

\begin{document}
\begin{abstract}
We pose the problem of transferring a Bose-Einstein Condensate (BEC) from one side of a double-well potential to the other as an optimal control problem for determining the time-dependent form of the potential. We derive a reduced dynamical system using a Galerkin truncation onto a finite set of eigenfunctions and find that including {\it three}  modes suffices to effectively control the full dynamics, described by the Gross-Pitaevskii model of BEC. The functional form of the control is reduced to finite dimensions by using another Galerkin-type method called the chopped random basis (CRAB) method, which is then optimized by a genetic algorithm called differential evolution (DE).  Finally, we discuss the extent to which the reduction-based optimal control strategy can be refined by means of including more modes in the Galerkin reduction.

\end{abstract}
\maketitle

\section{Introduction}
The dynamics of solitary waves in dispersive media with external potentials is a topic of widespread scientific interest, as it arises in many areas of application. For instance, in Bose-Einstein  Condensates (BEC) of, e.g., alkali gases, external potentials may be created using a variety of physical mechanisms including optical and magnetic fields, and may consist of one or a few wells or a periodic array, and may effectively confine the BEC to one, two, or three space dimensions~\cite{siambook, morsch,becbook1,becbook2}. Another appealing experimental setting is the nonlinear propagation of light through photonic crystals, and in the quasi-discrete realm of optical waveguides~\cite{joan,kivshar}. Here, a spatially-dependent index of refraction induces an effective potential~\cite{LEDERER20081}.

In both these applications, the simplest potential 
enabling bifurcation phenomena and nontrivial dynamics is arguably the double well potential. It has been studied intensely in the atomic
realm, following the hallmark theoretical work of~\cite{raghavan}. These predicted Josephson oscillations between the wells and quantum self-trapping were subsequently realized experimentally in~\cite{markus1}, as well as a dynamical symmetry-breaking bifurcation later observed in~\cite{zibold}.
More recent experiments have added damping and driving, which 
may present novel phenomena including stochastic resonance~\cite{markus3}.
Relevant double-well experiments have been conducted in the optical setting as well. The work
of~\cite{HaeltermannPRL02} considered the double-well potential in the context of 
twin-core self-guided laser
beams in Kerr media, while~\cite{zhigang} probed two-well dynamics using 
 photorefractive crystals. In this latter setting, additional phenomena were demonstrated in a \emph{triple-well} potential~\cite{kapitula}.

Naturally, this large volume of experimental developments and control has motivated a wide range
of theoretical explorations in numerous further directions. 
The relevant list is too long to do it proper justice, but
we mention some related studies.
Some more mathematical examples include the analysis of the double-well
bifurcation structure~\cite{AFGST:02, JW:bifdp}, 
the low-dimensional representation of the associated dynamical
problem (and its fidelity)~\cite{Kirr,MW10}, and the effect of changing the nonlinear exponent  on the bifurcation~\cite{sacchetti,KKP11}.
Among the many more physical examples are the interactions of multiple dispersive
(e.g., atomic) species~\cite{WANG20082922,gunay,tian}, incorporating
beyond-mean-field (i.e., many-body) effects~\cite{masiello,polls},
and the effect of larger spatial dimensions (and possibly
four wells)~\cite{PhysRevE.80.046611}, among others.

In this work, we  aim to apply the deep understanding of the
existence, stability and nonlinear dynamics to the context
of optimal control of BEC~\cite{kirk2004optimal,brif2010control}. The latter
methodology has long been recognized as a versatile
tool for engineering on-demand quantum states of interest. 
In the early stages, the framework of magnetic microtraps
controllable by external parameters including radio-frequency
fields or/and wire currents was used to enable the preparation
of desired states~\cite{Hohenester} (see also the detailed analysis
of the relevant methodologies and their numerical \Matlab-based 
implementation in~\cite{Hohen2,Hohen3}). Subsequently, such ideas have
been applied to fully three-dimensional
settings, e.g., in the work of~\cite{Mennemann} and have been
used recently by a subset of the present authors in order to 
re-orient the density distribution of an atomic BEC and alter the topology of its support~\cite{me}.

Here, more concretely, we intend to show how the low-dimensional
representation available in the context of double (and more
generally few~\cite{goodman,PhysRevE.80.046611}) well potentials
can be used as a basis for performing optimal control analysis
and for achieving desired end quantum states, both in the low-dimensional
setting, but also in the full mean-field model
of the Gross-Pitaevskii (nonlinear Schrödinger-type) partial
differential equation (PDE)~\cite{becbook1,becbook2}. Along the way, we learn
the following lesson, which we find interesting and important. 
The two-mode expansion is prevalent in the study of the double-well system,
and widely acknowledged to describe the dynamics, both qualitatively and even 
quantitatively. By contrast, we find that in the context of optimal control, we must 
include (at least) the third mode in the expansion to achieve useful agreement.
We believe that such lessons may prove useful for other
practitioners in related contexts.

Our presentation is structured as follows. In Section~\ref{section:Two}, we  present the physical
and mathematical setup and its reduced (two- and three-mode)
representation. Subsequently, in Section~\ref{section:OC}, we  present the proposed optimal control strategy. In Section~\ref{section:results}, we 
display numerical results. Finally, Section~\ref{section:conclusions} provides a summary of our conclusions, as well as a number of directions for future study.


\section{Derivation of Reduced Model Systems}\label{section:Two}
The approach to optimization we propose here is to apply optimal control to a finite-dimensional model system, whose derivation we outline in this section. In particular, we use Galerkin truncation to derive a low-dimensional Hamiltonian system whose dynamics capture the essence of the full dynamics. The latter, in turn, 
is described by a Gross-Pitaevskii equation (GPE) in one spatial dimension. This system may be derived from a three-dimensional model in the presence of an anisotropic potential that squeezes the condensate into an effectively one-dimensional arrangement; see details and nondimensionalization, e.g., in~\cite{siambook}. The initial and (final) desired conditions used throughout are set to the stable fixed points of the finite-dimensional model Hamiltonian we have derived. These fixed points correspond to the two asymmetric states which exist in the presence of the barrier. Indeed, our aim is to drive the atomic mass from a state predominantly localized in one well to a state  localized in the other. The efficiency will be determined on the basis of how successful such a transfer is with an appropriate definition of ``fidelity''  to the intended end state;
more detail will follow the analysis.

The GPE model is a nonlinear Schrödinger equation with a spatial potential
\begin{equation}\label{eq:GPE}
    i \partial_t \psi = 
    \mathcal{L}\left( x,\vw(t)\right)\psi + \mathcal{N}(\psi) =
    -\frac{1}{2}\partial_x^2\psi+V\left(x,\vw(t)\right)\psi+\abs*{\psi}^2\psi,
\end{equation}
where $\vw(t)$ is a time-dependent vector of $C^0$ control functions. In addition to conserving a Hamiltonian energy, this system conserves the mass
\begin{equation}
M = \int_{-\infty}^{\infty} \abs*{\psi}^2 \dd x,
\label{eq:mass}
\end{equation}
which in the BEC context is interpreted as the total number of atoms in the condensate. The potential $V$ is chosen as the superposition of a quadratic 
confining potential, usually implemented via  magnetic fields~\cite{becbook1,becbook2}, and a thin, tall barrier at the center, 
typically induced by an optical beam~\cite{zibold}. Added together, these form a prototypical
double-well potential; see, also,~\cite{PhysRevE.74.056608}. In particular, the potential takes the form
\begin{equation}
    V=\frac{1}{2}u(t)x^2+v(t)\delta(x),
\end{equation}
where the first term models the magnetic confinement and the second term models the localized repulsive barrier at the center, as we $v(t)>0$. The time-dependent parameter vector is thus given by $\vw(t)=(u(t),v(t))^T.$ 

Let $\Phi_{\vw} = \left\{ \varphi_n(x;\vw)\in L^2(\RR) \ |\  n=0,1,\ldots \right\}$ be the set of normalized eigenfunctions of the linear Schrödinger eigenvalue problem
\begin{equation}\label{eq:lineig}
    \mathcal{L}(x,\vw)\varphi_n=E_n\varphi_n,
\end{equation}
for a fixed parameter vector $\vw$.  Because $\Phi_{\vw}$ is complete in $L^2(\RR)$, we may represent the solution to Eq.~\eqref{eq:GPE} at time $t$ by the infinite series
$$
\psi(x,t)  =\sum_{n=0}^{\infty}c_n(t)\varphi_n(x;\vw(t)).
$$
Plugging this representation into Eq.~\eqref{eq:GPE}, and projecting both sides of the equation onto $\varphi_n$---in the $L^2(\CC)$ sense---yields an evolution equation for $c_n(t)$. Together, the evolution of the infinite vector $\vb{c} = (c_0(t), c_1(t),\ldots)$ of complex amplitudes is then equivalent to the evolution of $\psi$ under GPE.

To derive an approximate reduced system, we consider the truncated series superposition of instantaneous eigenfunctions
\begin{equation}\label{eq:GPEGal}
    \psi_{N+1}^{\rm Gal}:=\sum_{n=0}^{N}c_n(t)\varphi_n(x;\vw(t))
\end{equation}
for some fixed value of $N<\infty$. Ignoring any truncation error due to the terms in the omitted tail of the series yields the system of interest. Such a truncation has been rigorously justified in certain very simple cases, e.g.~\cite{MW10,gmw_2015}, but the method is commonly applied without rigorous justification.

\subsection{Two-Mode Expansion}\label{section:2m}
By setting $N=1$ in Expansion~\eqref{eq:GPEGal},  we find the following two-mode 
Hamiltonian system~\cite{PhysRevE.74.056608}:
\begin{equation}\label{eq:2msys}
\begin{split}
    i\dot{c}_0&=\frac{\partial\mathcal{H}}{\partial\bar{c}_0}=\alpha c_0+\gamma_0 \abs{c_0}^2c_0+\gamma_2 \left(c_1^2 \bar{c}_0+2 c_0\abs{c_1}^2\right),\\
    i\dot{c}_1&=\frac{\partial\mathcal{H}}{\partial\bar{c}_1}=\beta c_1+\gamma_1 \abs{c_1}^2c_1+\gamma_2 \left(c_0^2 \bar{c}_1+2 c_1\abs{c_0}^2\right),
\end{split}
\end{equation}
with instantaneous projection coefficients given by
\begin{align} \label{2modeconstants}
&\alpha=\ip{\mathcal{L}\varphi_0}{\varphi_0},\
  \beta=\ip{\mathcal{L}\varphi_1}{\varphi_1},\
  \gamma_0=\norm*{\varphi_0}^4,\ 
  \gamma_1=\norm*{\varphi_1}^4,\
  \gamma_2=\ip{\varphi_0^2}{\varphi_1^2}.\ 
\end{align}
Its Hamiltonian reads:
\begin{equation}
    \mathcal{H}=\alpha\abs{c_0}^2+\beta\abs{c_1}^2+\frac{\gamma_0}{2}\abs{c_0}^4+\frac{\gamma_1}{2}\abs{c_1}^4+\gamma_2\left(\Re\left\{c_0^2\bar{c}_1^2\right\}+2\abs{c_0}^2\abs{c_1}^2\right).
\end{equation}
This expansion holds under the general assumption that $V(x,\vw(t))= V(-x,\vw(t))$, i.e., that the potential is even. This system conserves a discrete form of the mass defined in Eq.~\eqref{eq:mass},
\begin{equation}\label{eq:2Md}
    \Md(t)=\abs*{c_0(t)}^2+\abs*{c_1(t)}^2.
\end{equation} 

This system has stationary solutions of the form $(c_0(t),c_1(t))= (\rho_0,\rho_1)e^{-i\Omega t}$. In particular, it has a solution corresponding to the nonlinear continuation of the ground state with $\rho_1=0$ and $\Omega = \alpha + \gamma_0 \rho_0^2$, and a second solution corresponding to the nonlinear continuation of the excited state, with $\rho_0=0$ and $\Omega = \beta+ \gamma_1 \rho_1^2$. 
In the absence of a barrier, i.e., for $v=0$, and total mass $\Md=1$, these are the only such states, and both are linearly stable. 
In the presence of a barrier, however, the excited state can become unstable in a symmetry-breaking pitchfork bifurcation.

We may take advantage of the conservation law 
of Eq.~(\ref{eq:2Md}) to reduce the system from two degrees of freedom to one as follows, which provides a convenient visualization of the  dynamics and bifurcation. Consider the canonical transformation
\begin{equation}\label{eq:firstcan}
c_0=Ae^{i\theta},\quad c_1=(q+ip)e^{i\theta}.
\end{equation}
We reduce the number of degrees of freedom from two to one using the conserved mass~\eqref{eq:2Md} which now reads $A^2=\Md-q^2-p^2$. The Hamiltonian in these coordinates is given by
\begin{align}\label{eq:qpHam}
    \mathcal{H}=\frac{\gamma_0 \Md^2}{2}+\alpha  \Md+q^2 \left(\beta -\alpha +\Md\left(3\gamma_2 -\gamma_0\right)\right)+p^2 \left(\beta -\alpha +\Md\left(\gamma_2 -\gamma_0\right)\right)  \nonumber \\
    +\left(\frac{\gamma_0}{2}+\frac{\gamma_1}{2}-\gamma_2\right) p^4+\left(\gamma_0+\gamma_1-4 \gamma_2\right) p^2 q^2+\left(\frac{\gamma_0}{2}+\frac{\gamma_1}{2}-3 \gamma_2\right) q^4.
\end{align}

\begin{figure}[htbp]
\begin{centering}
\subfigure{\includegraphics[width=0.45\textwidth]{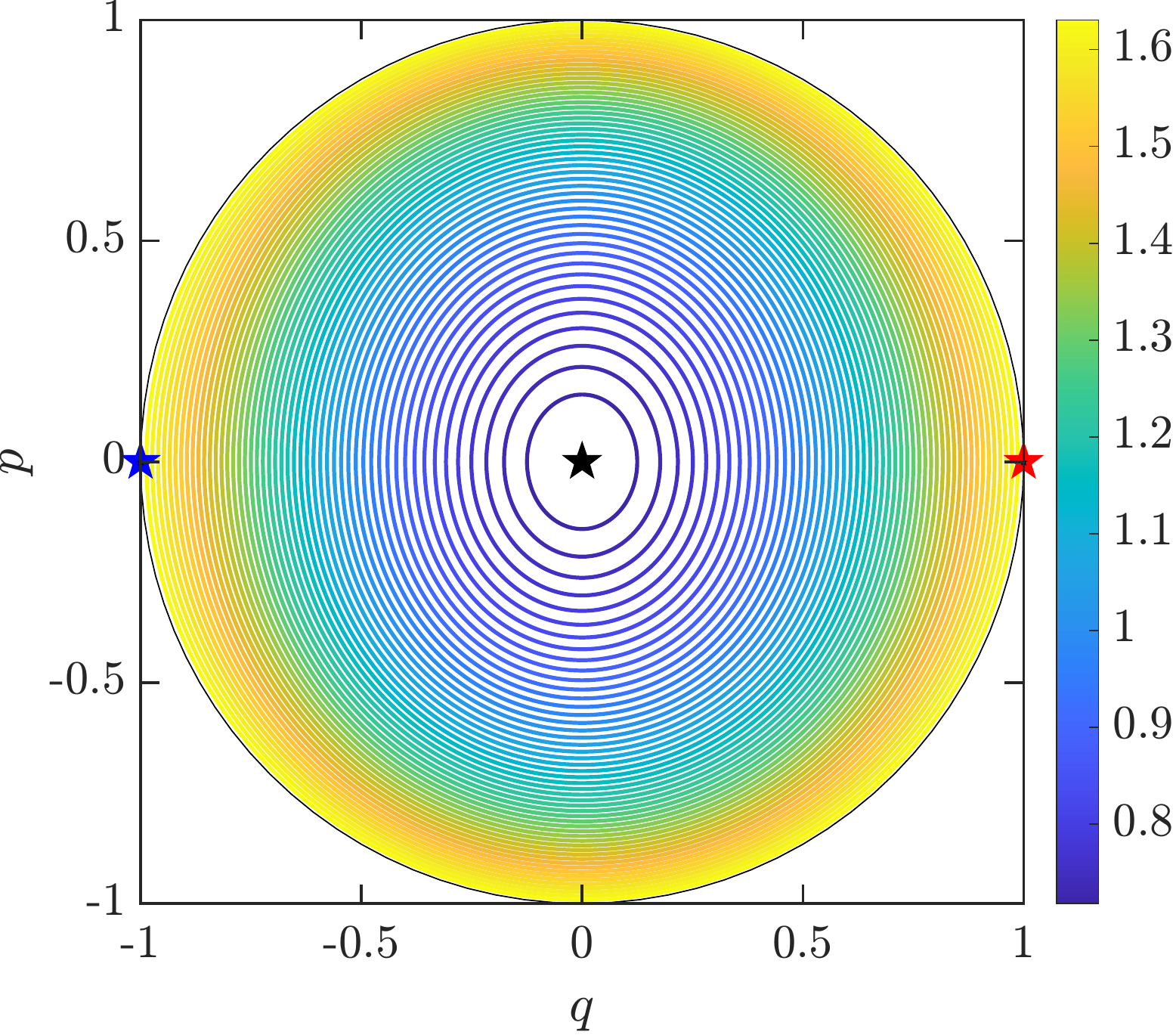}}
\subfigure{\includegraphics[width=0.45\textwidth]{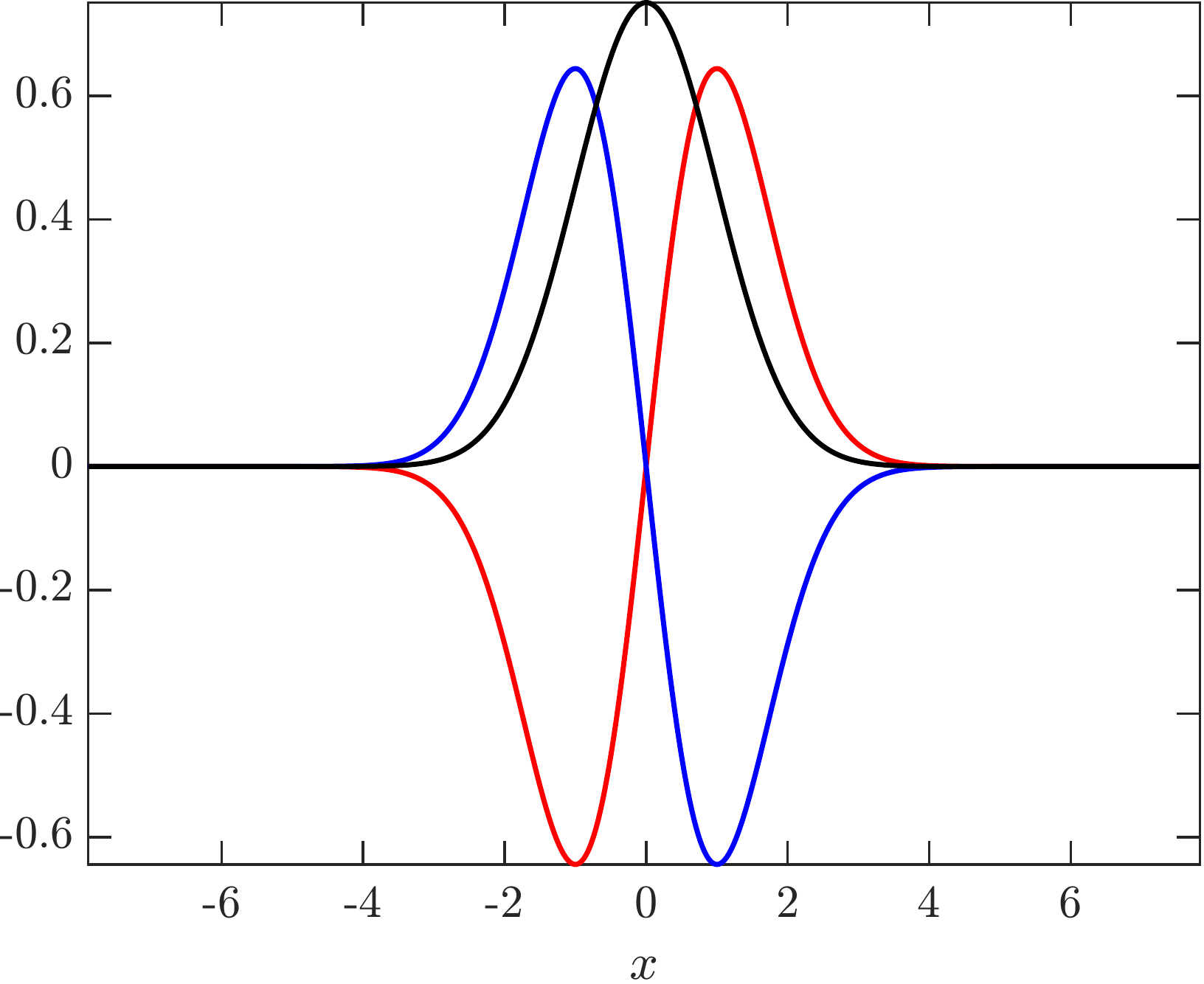}}
\subfigure{\includegraphics[width=0.45\textwidth]{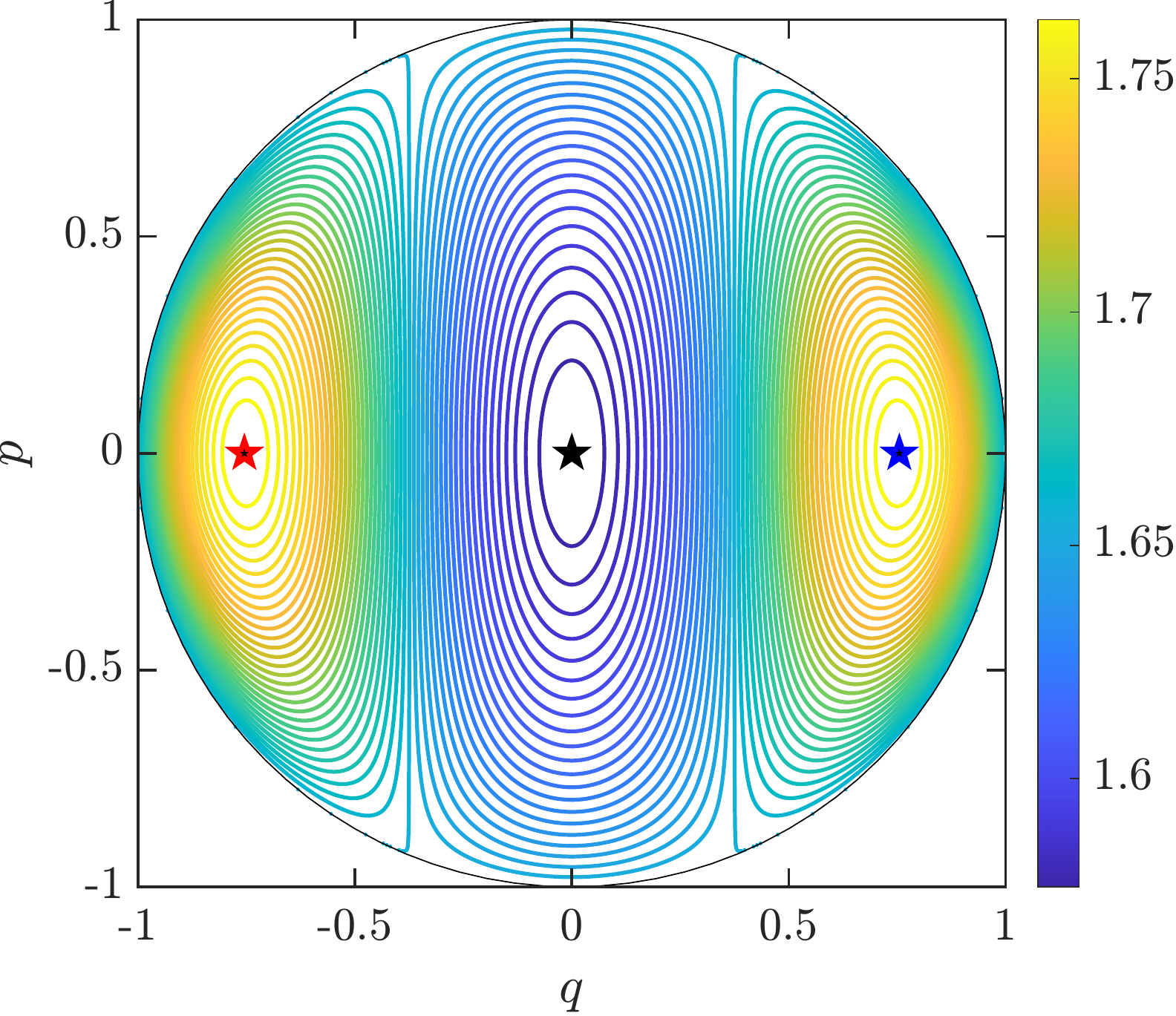}}
\subfigure{\includegraphics[width=0.45\textwidth]{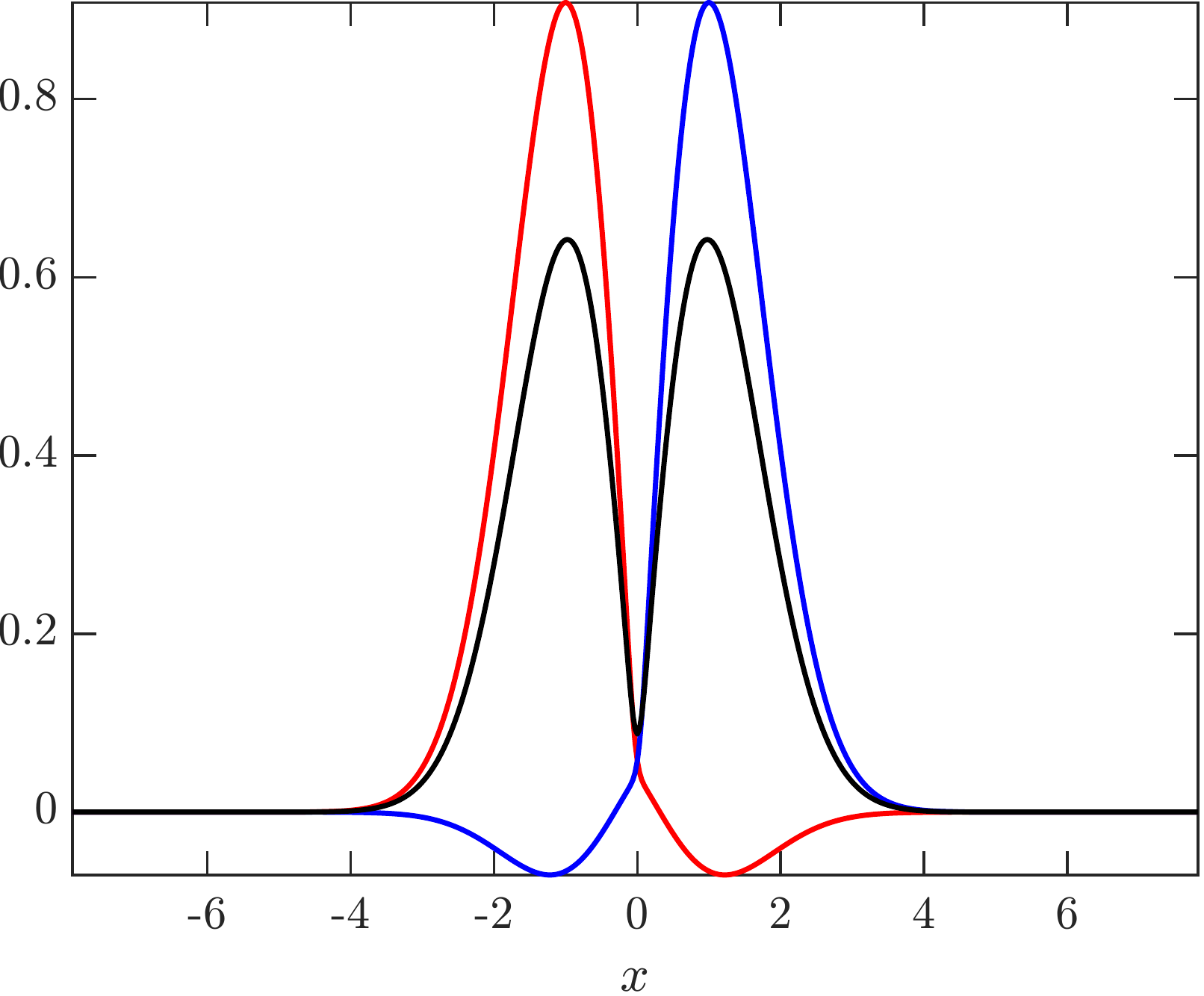}}

\caption{Left panels: the phase space of system~\eqref{eq:qpHam} for $v=0$ (top) and $v=10$ (bottom), with the ground state marked with a black star, showing that two fixed points have bifurcated from the bounding circle as $v$ was raised, now sitting at approximately $(q,p)=(\pm0.7550,0)$. Right panels, same parameter values, show in black the ground state standing wave and in red/blue the odd-symmetric standing waves (top) and the symmetry broken standing waves $\psi_0$ and $\psid$ (bottom).
}\label{fig:PS1}
\end{centering}
\end{figure}

We show the phase portraits associated with Hamiltonian~\eqref{eq:qpHam} for values of $v=0$ and $v=10$ with fixed $\Md=1$ with $u=1$ in the first column of Figure~\ref{fig:PS1}. In the reduced system, the ground state standing wave becomes a fixed point at the origin, and the excited state standing wave becomes the boundary circle $p^2+q^2=1$. For $v=10$, two new asymmetric states have emerged from the odd solution and appear as fixed points on the $q$-axis.  The right column shows the  standing waves constructed from the Galerkin ansatz, including the initial and desired states of the control problem $\psi_0$ and $\psid$.

For $v=0$, the double-well structure is absent (i.e., the
the setting is one of a parabolic trap with equidistant linear eigenvalues),
hence the two-mode reduction is not expected to provide an adequate representation
of the dynamics (except for very low masses $\Md$).
Of course, to find the coefficients $c_0$ and $c_1$, we back substitute using the above canonical transformations through
\begin{equation} \label{eq:polar}
c_0=\sqrt{\Md-q^2-p^2}e^{i\theta},\quad c_1=(q+ip)e^{i\theta},
\end{equation}
where $\theta\in[0,2\pi]$ is arbitrary by the phase invariance of the Hamiltonian~\eqref{eq:qpHam}. Without loss of generality, we choose $\theta=0$ so that $\psi_2^{\rm Gal}$, given by Equation~\eqref{eq:GPEGal}, is real. 

Hereafter, we fix the barrier height to $v=10$, and choose the value of $\Md$ such that the stable fixed points with $q\neq0$ are asymmetric states, as shown in Figure~\ref{fig:PS1}. 
That is, we operate within the symmetry-broken regime of the double-well
potential. We note that for any finite-strength barrier, the stable state is "partially fragmented" in that a nonzero fraction of the mass resides in each well; see~\cite{PhysRevA.59.3868} for a theoretical account within the many-body formalism of BEC. As the value of $v$ is further increased, the fragmentation is lessened, which provides an obvious strategy for mitigating fragmentation.

Moving forward, the phase space shown in (the bottom panels of) Figure~\ref{fig:PS1} makes clear the goal of our optimal control problem: to find functions $u(t)$ and $v(t)$, with fixed and identical initial and terminal conditions, that drive the system from one asymmetric steady state $\psi_0(x)$ to the other one $\psid(x)$. 
From a physical perspective, our aim is to drive atoms from a state in which most reside in one well, into one in which most reside in the other, using the experimentally-developed
ability to temporally drive double-well potentials~\cite{markus3} and
more specifically magnetic and optical confining beams~\cite{Hohenester,becbook1,becbook2}.
This control problem is  mathematically formulated in Section~\ref{section:OC}.

\subsubsection*{Numerical Validation}
Before describing the optimal control problem in detail, we numerically test the ability of the two-mode system~\eqref{eq:2msys} to approximate the dynamics of the GPE~\eqref{eq:GPE} with appropriate initial conditions. We consider the evolution of 
the initial condition  $\psi(x,0) = \varphi_0(x)$ as shown in Figure~\ref{fig:PS1} subject to the GPE with imposed controls
\begin{equation}\label{eq:trialcont}
    u_{\rm trial}(t)=1+2\sin\left(\frac{\pi t}{T}\right) \qand
    v_{\rm trial}(t)=10\cos^4\left(\frac{\pi t}{T}\right)
\end{equation}
over the time interval $t\in[0,T]$, with $T=2.$  We have proposed these trial controls based on ad hoc reasoning, using on the following 
partial intuition: before performing an optimization, we suspect that
an optimal potential would allow the barrier at the origin to lower so that the initial mass on the one well can be transferred  significantly to the
other well, followed by a raising of the barrier anew in order to localize the wavefunction into the desired well of the potential. 
The choices in the modulation of the parabolic trap, the strength of the 
localized barrier, and the length of time in the simulation were all chosen arbitrarily in this dynamical example.

It does not escape us that in physical settings involving quasi-1d double wells
in atomic BECs, the width parameter $u$ is constrained to be $u(t) \ll 1$ for the
quasi-1d reduction to hold. We have considered such scenarios as well, finding
qualitatively similar results, as regards the optimal control framework discussed later on in Section~\ref{section:OC} within the Galerkin truncation, although over considerably longer time scales.

We numerically integrate the two-mode system~\eqref{eq:2msys}  using \Matlab's \texttt{ode45}. Its time-dependent coefficients from Eq.~\eqref{2modeconstants} depend on $\vw(t)$  through the instantaneous eigenfunctions $\varphi_0$ and $\varphi_1$. While closed form expressions for these eigenfunctions are determined, for each value of $\vw$, in terms of hypergeometric functions~\cite{Viana-Gomes2011}, we find it simpler to solve the associated eigenproblem numerically at each time step using \Matlab's $\texttt{eig}$ command.
Indeed, while the former possibility is particular to the potential considered
herein, the latter can be extended to arbitrary time-dependent potentials.

We solve the GPE~\eqref{eq:GPE} using a second-order Fourier split-step method and approximate the delta function by a narrow Gaussian
\begin{equation}
    \delta(x)=\lim_{a\to\infty}\frac{a}{\sqrt{\pi}}e^{-a^2{x^2}},
\end{equation}
with $a=12$ here and in all subsequent computations. Although the Fourier split-step method we use to solve the GPE is quite standard, we provide details about the method in~\ref{section:GPENum} for completeness. Throughout this work, we find that a uniform spatial discretization on the truncated interval $x\in[-5\pi,5\pi]$ of $2^{11}$ points and a temporal discretization of $T/h$ points, choosing $h=2^{-8},$ yields an accurate and stable computation of the GPE dynamics.

In what follows, we need a way to measure the agreement between the solutions to GPE and the finite-dimensional approximation defined by expansion~\eqref{eq:GPEGal}. We define the projected wavefunction as
 \begin{equation}\label{eq:psiproj}
     \psi^{\rm proj}_{N+1}=\sum_{n=0}^N\left<\psi^{\rm GPE}(\cdot,t),\varphi_n(\cdot,u(t))\right>\varphi_n(x,u(t)),
 \end{equation}
where $\psi^{\rm GPE}$ solves Equation~\eqref{eq:GPE}. We quantify an expected upper bound on the Galerkin approximation~\eqref{eq:GPEGal} through the relative error
\begin{equation}\label{eq:relerr}
    \mathcal{E}_{N+1}(t) = 
    \frac{\norm*{\psi^{\rm GPE}-\psi_{N+1}^{\rm proj}}^2}{M} = 
    \frac{\left\lVert\sum_{n=N+1}^\infty \ip{\psi^{\rm GPE}(\cdot,t)}{\varphi_n(\cdot,\vw(t))}\varphi_n(x,\vw(t))\right\rVert^2}{M}, 
\end{equation} 
where $M$ is defined by Eq.~\eqref{eq:mass} and the norm is taken in $L^2(\CC)$. The second equality, which is interpreted as the relative mass content which has been excited beyond the few-mode representation at order $N$, relies on the fact that the GPE~\eqref{eq:GPE} is well-approximated by Formula~\eqref{eq:psiproj} in $L^2\left(\CC\right)$, i.e., $\mathcal{E}_{N+1}\to0$ as $N\to\infty.$
Therefore, we rely on various comparisons among Equations~\eqref{eq:GPEGal},~\eqref{eq:psiproj}, and~\eqref{eq:relerr} when discussing the extent to which few-mode representations effectively shadow the full-dynamical picture given by the GPE~\eqref{eq:GPE}. Also, note that although wavefunctions are normalized to have unit mass throughout this work, we include $M$ in the above definition of $\mathcal{E}_{N+1}$ to maintain clarity of how the analysis should be performed given a different scaling of Equation~\eqref{eq:GPE}.

\begin{figure}[htbp]
\begin{centering}
\subfigure{\includegraphics[width=0.45\textwidth]{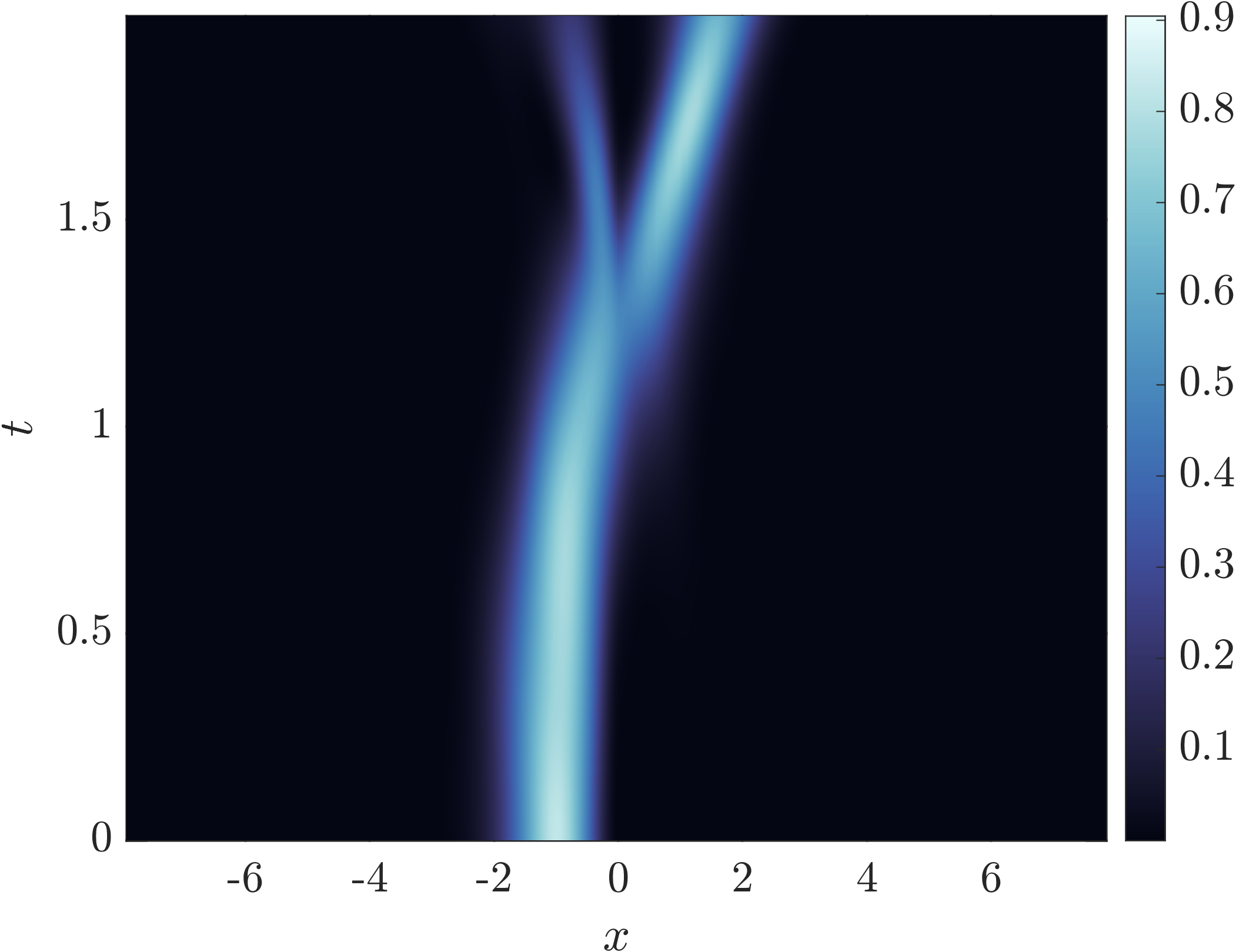}}
\subfigure{\includegraphics[width=0.45\textwidth]{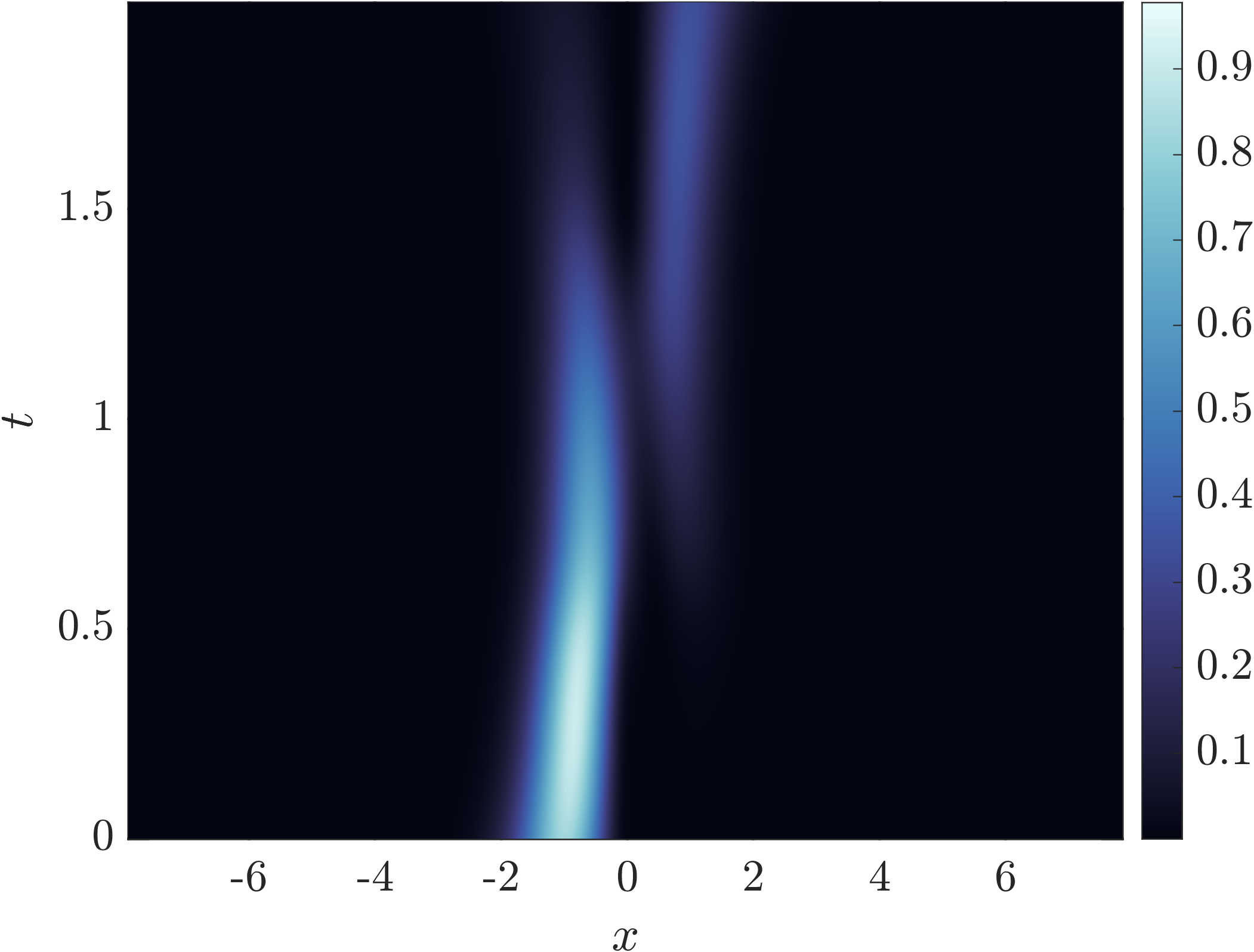}}
\subfigure{\includegraphics[width=0.45\textwidth]{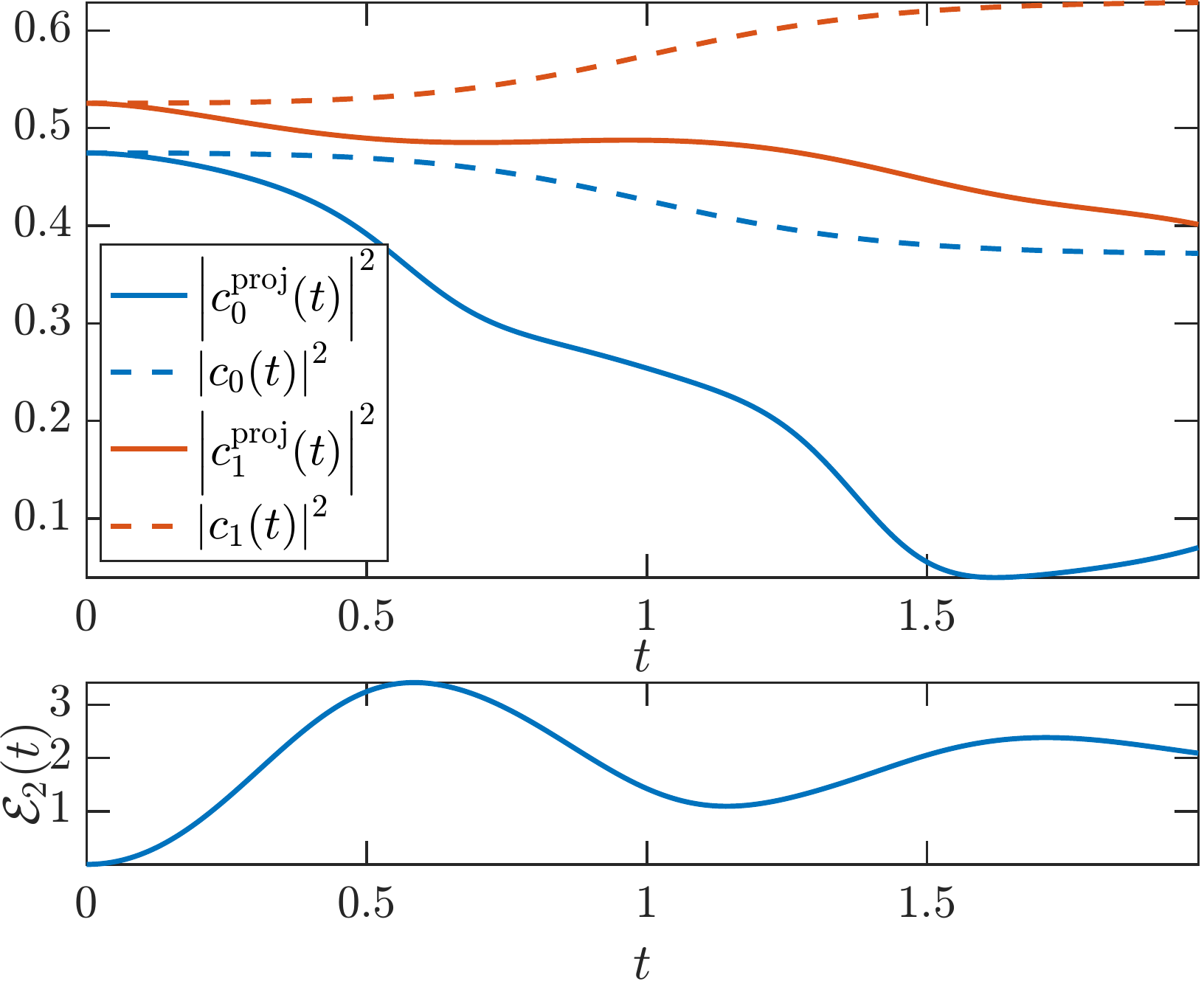}}
\subfigure{\includegraphics[width=0.45\textwidth]{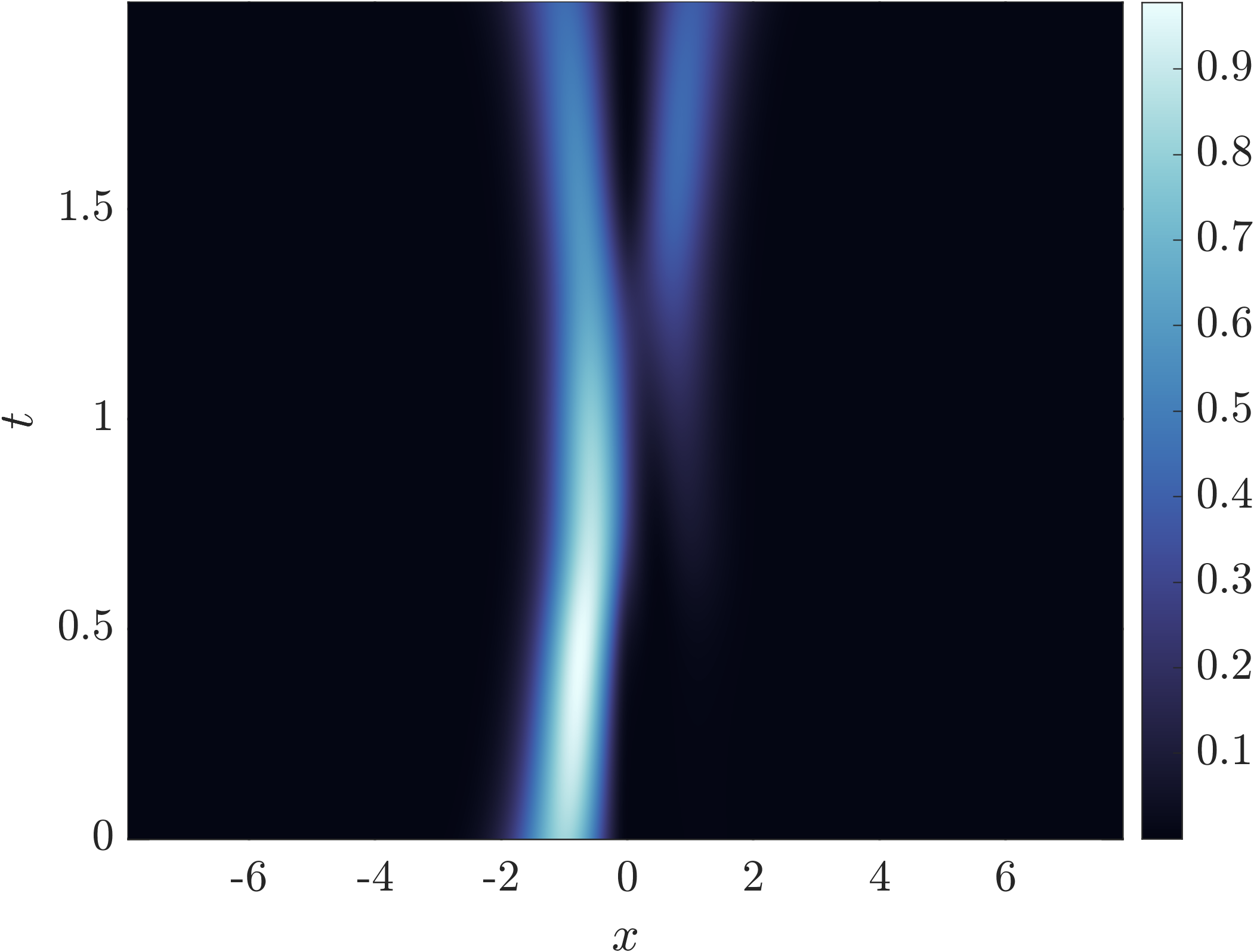}}
\caption{A comparison between numerical solutions to the GPE~\eqref{eq:GPE} and the two-mode system given by Equations~\eqref{eq:2msys}. Top left: a full numerical simulation of the GPE~\eqref{eq:GPE} with the colormap showing the  squared amplitude. Top right: the projected wavefunction $\psi_2^{\rm proj},$ in absolute value squared, as in Equation~\eqref{eq:psiproj}. Bottom left: a comparison of numerically computed and projected coefficients $c_n(t)$ for $n=0$
and $n=1$, as well as the loss of mass to higher modes given by the error formula~\eqref{eq:relerr}. 
Bottom right: the wavefunction $\psi_2^{\rm Gal}$, as defined by Equation~\eqref{eq:GPEGal} and also in absolute value squared, with coefficients shown in the bottom left panel. 
}\label{fig:2mT}
\end{centering}
\end{figure}

With those preliminaries, we are ready to show the results of the simulations.
Figure~\ref{fig:2mT} shows simulations the GPE~\eqref{eq:GPE} and the two-mode model system~\eqref{eq:2msys}. It presents three colormaps: the wave function $\psi^{\rm GPE}$, the projection of the wave function onto the first two instantaneous eigenfunctions, i.e., $\psi^{\rm Gal}_2$, and the wave function $\psi_{N+1}^{\rm Gal}$ constructed from the solution to the two-mode model~\eqref{eq:2msys} using Formula~\eqref{eq:GPEGal}.
A fourth plot shows the solutions of Equation~\eqref{eq:2msys} as well as the  coefficients defining the projected wavefunction~\eqref{eq:psiproj} from the GPE solution.

We make two observations based on these plots. First, our naively chosen control functions~\eqref{eq:trialcont} crudely transfer the bulk of the solution from the left potential well to the right, in both the GPE system and the two-mode model system. Second, and despite the first observation, the agreement between the two dynamics is poor. This is seen in the poor agreement to the computed values of the coefficients and through the large relative error $\mathcal{E}_2$(t). 

This is a central observation of this study: while the two-mode reduction
effectively describes the bifurcation structure and the dynamical
evolution in the vicinity of the symmetric and asymmetric equilibrium. However, for the more highly non-equilibrium transfer proposed herein, a  representation requiring more modes becomes necessary. In that light, we now pursue a three-mode reduction of the system.

\subsection{Three-Mode Expansion}

In order to increase the fidelity of the reduced model to the full GPE, we now compute a model equation using $N=2$ in the expansion~\eqref{eq:GPEGal}. The form of the Hamiltonian system is long and fairly unenlightening, so we display the associated Hamiltonian~\eqref{eq:3mHam} in Appendix~\ref{section:3m}. To simplify the search for stationary solutions, we perform a canonical transformation similar to Eq.~\eqref{eq:polar}, yielding  Hamiltonian~\eqref{eq:3mPolHam}. Not surprisingly, we find asymmetric standing waves similar to $\psi_0$ and $\psid$ from Figure~\ref{fig:PS1}. In fact, the contribution to these modes from $c_2$ is no greater than one part in $10^{-10}$ in absolute-value squared. This numerically justifies the simplification of using the fixed point, corresponding to the left-asymmetric state from Figure~\ref{fig:PS1}, with no contribution from the third mode. Additionally, this also allows for consistency in testing the accuracy of the three-mode model.

\begin{figure}[htbp]
\begin{centering}
\subfigure{\includegraphics[width=0.45\textwidth]{3mT3.png}}
\subfigure{\includegraphics[width=0.45\textwidth]{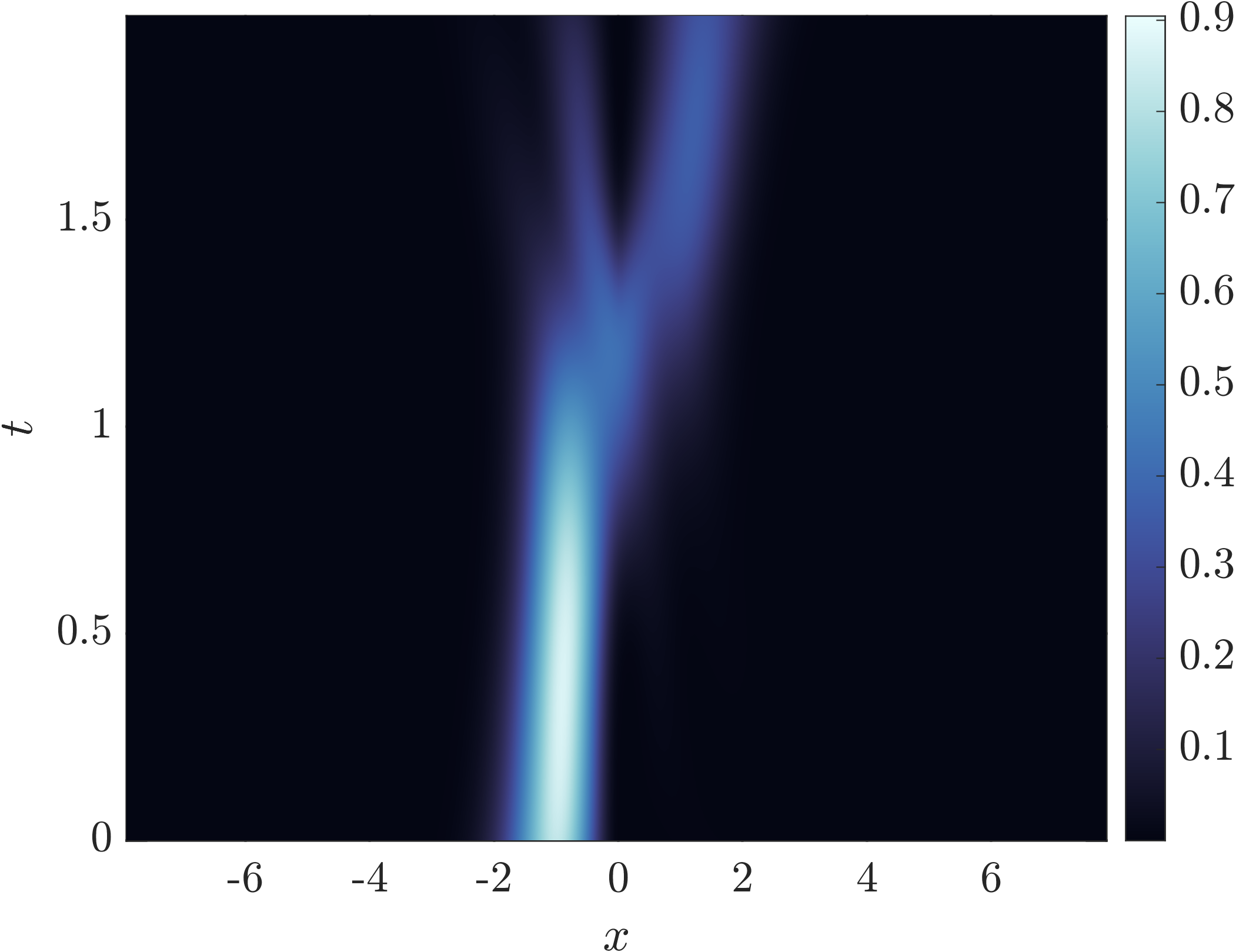}}
\subfigure{\includegraphics[width=0.45\textwidth]{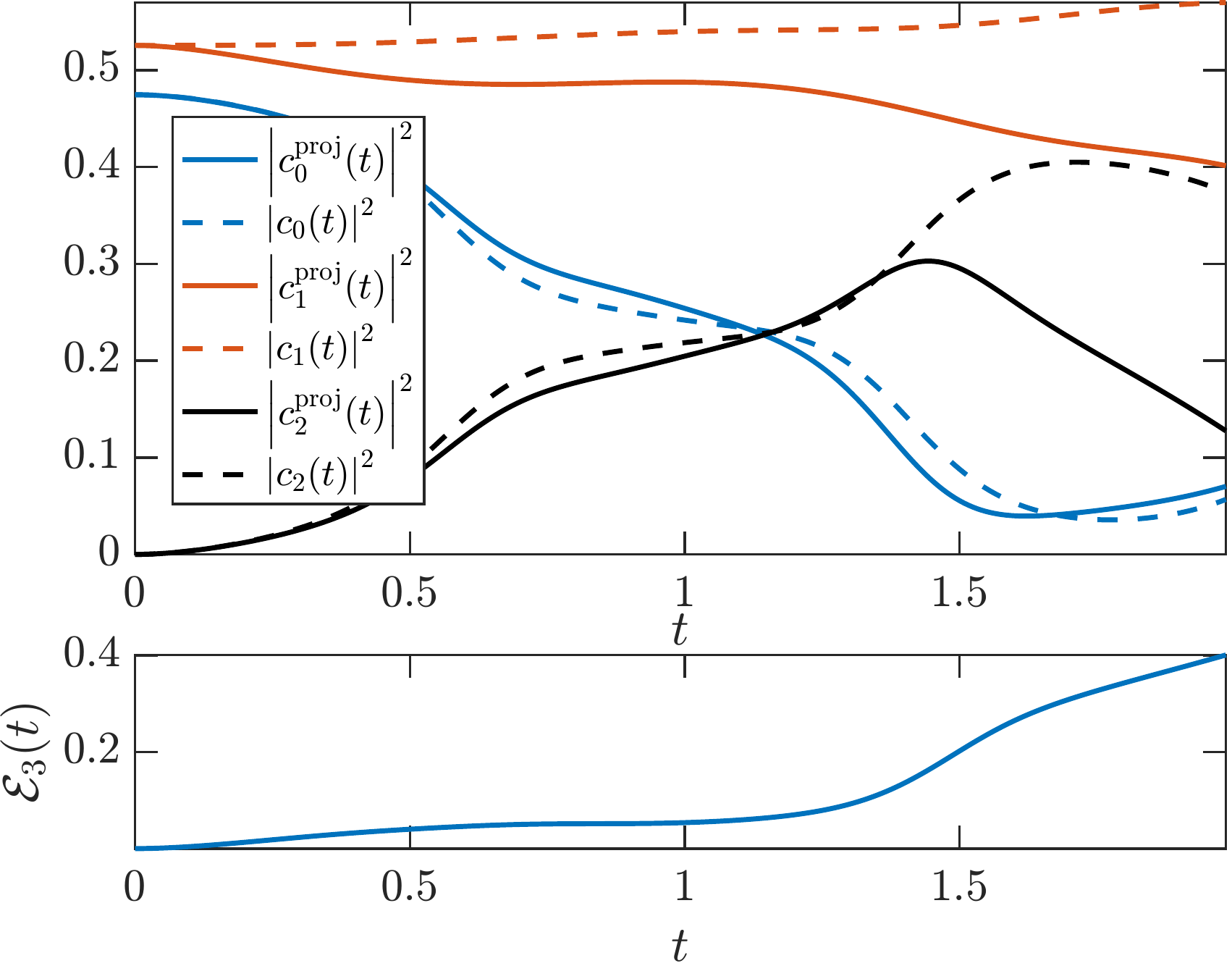}}
\subfigure{\includegraphics[width=0.45\textwidth]{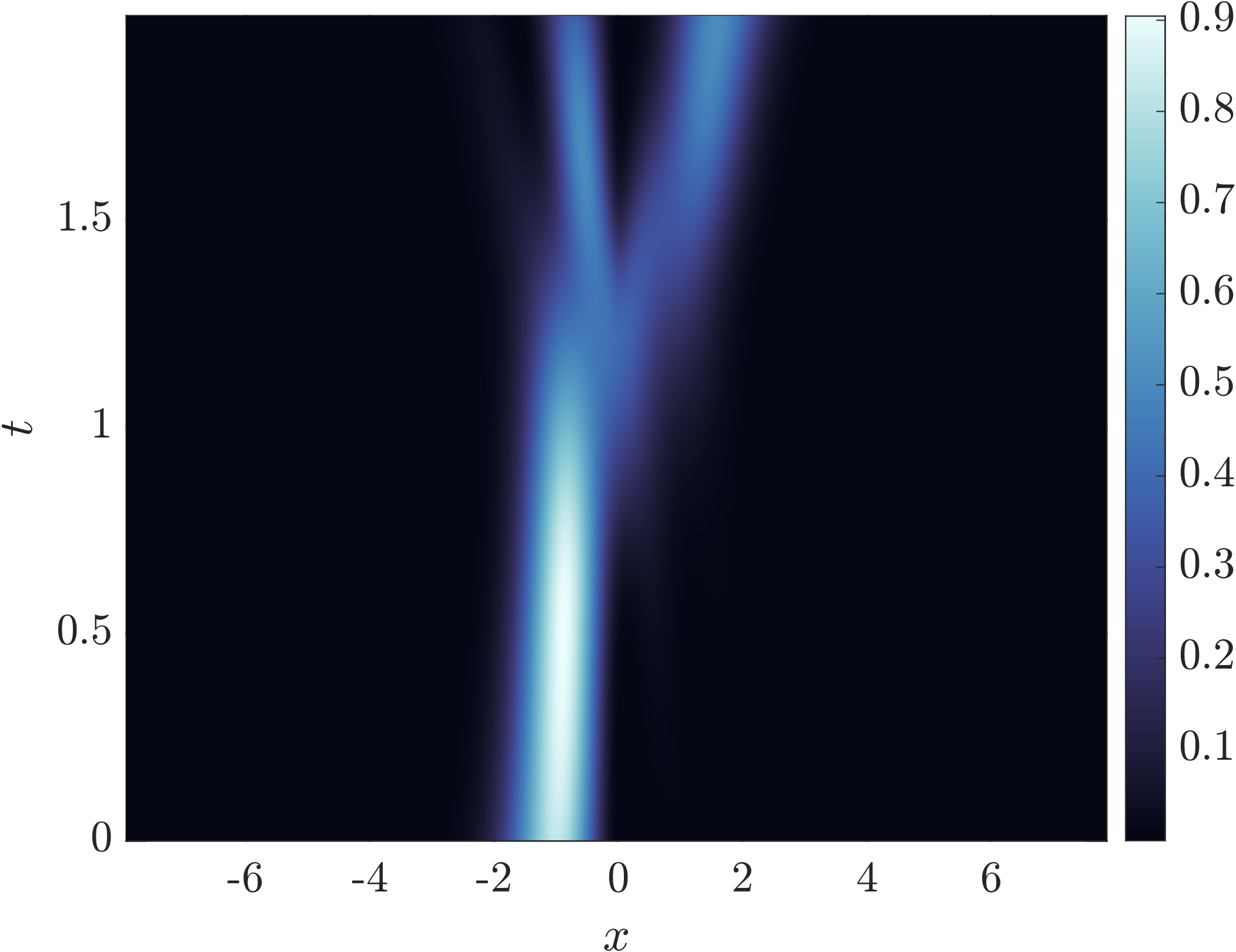}}
\caption{A comparison between numerical solutions to the GPE~\eqref{eq:GPE} and the three-mode system given by Hamiltonian~\eqref{eq:3mHam} as was similarly presented in Figure~\ref{fig:2mT}. Here, we see a substantial reduction in the error $\mathcal{E}_3(t)$ in contrast with the error $\mathcal{E}_2(t)$ shown in Figure~\ref{fig:2mT}, allowing the three-mode model to more effectively capture the full dynamics given by Equation~\eqref{eq:GPE}.}\label{fig:3mT}
\end{centering}
\end{figure}

Using the above-mentioned initial condition, we run the same test shown in Figure~\ref{fig:2mT}, but for the three-mode model, as shown in Fig.~\ref{fig:3mT}. In this case, the approximation is considerably more accurate, especially up to intermediate times. For example at $t=1$, the relative error $\mathcal{E}_3$ is about 0.05. More importantly, we see that the even modes in the reduction effectively capture the even-projected dynamics of the GPE~\eqref{eq:GPE} for up to intermediate times.

However, we lose a great deal of accuracy for the remainder of the simulation, where, by $t=T$ the relative error has grown to about 0.40. The plot of $c_1$ and $c_1^{\rm proj}$  shows that this is due to a significant excitation of higher-order odd modes which have been neglected. Despite this, we find that the three-mode model performs well enough in our pursuit of optimal controls. In Section~\ref{section:results}, we quantify the contribution of the higher modes to the relative error~\eqref{eq:relerr} while the condensate is being controlled in the full dynamical setting. We find that the inclusion of just one more odd mode substantially reduces the relative error. There, we discuss this in greater detail, and, for now, leave the pursuit of higher-dimensional models as a subject for future work.  

We make a final comment on the role that the mass $M$ has on the relative error $\mathcal{E}_3$. Indeed, for smaller values of the  mass $M$, we see a reduction in $\mathcal{E}_3$, as expected since nonlinear interactions in Equation~\eqref{eq:GPE} are substantially smaller. Using $M=0.2$ which is only slightly above the symmetry-breaking bifurcation value, we find $\mathcal{E}(T)$ to be near 0.25. Although this error has been reduced, enhancing the efficacy of the reduced dynamical system, the mode $\psi_0$ corresponding to the asymmetric stationary state of Hamiltonian~\eqref{eq:qpHam} is only slightly asymmetric, failing to meet our stated goal of concentrating the mass in a single well.  For this reason, we continue to use $M=1$ in what follows, since this strikes a good balance between a small enough relative error and an adequate initial condition $\psi_0$. 

\section{Optimal Control Strategy}\label{section:OC}

The optimal control problem we pursue, motivated by the previous section, is to construct a function $\vw(t)$ that drives an initial state $\vb{c}_0$ to the desired state $\vb{c}_{\rm d}$ under the dynamics of the two- or  three-mode model. More precisely, the control problem is to find a local minimizer of the problem
\begin{equation}\label{eq:OCP}
    \min_{\vw(t)\in\mathcal{W}}\mathcal{J} =
    \min_{\vw\in\mathcal{W}}\left\{M^2_{\rm d}-\abs*{\ip{\vb{c}_{\rm d}}{\vb{c}_T}}^2\right\},
\end{equation}
where the admissible space of controls is given by $\mathcal{W}=\{\vw(t) \in C^0([0,T]):\vw(0)=\vw(T)=\vw_b\}$, for a prescribed value of  $\vw_b$. We call this objective functional the discrete infidelity, as opposed to the full infidelity
\begin{equation}
    \mathcal{J}_{\rm full}=M^2-\abs*{\ip{\psid}{\psi_T}}^2
\end{equation}
first used by Hohenester, et al.~\cite{Hohenester}. In both cases, the infidelity penalizes misalignments of the final computed state with the desired state. 
Note that since the discrete mass $\Md$ is conserved by the dynamics governing $\vb{c}$, the optimal, perhaps unachievable, final state is the desired, which yields an optimal infidelity of zero.

An advantage of the infidelity~\eqref{eq:OCP} 
is that 
it is insensitive to the global phase of the dynamics, which is physically unimportant. A more traditional least-squares approach can be introduced via the objective
\begin{equation}\label{eq:LSQ}
    \mathcal{J}_{\rm d}^{\rm LSQ}=\frac{1}{M_{\rm d}}\abs*{\vb{c}_{\rm d}-e^{iS}\vb{c}_T}^2,
\end{equation}
where the phase $S\in[0,2\pi].$ Although we do not pursue such an optimization we report the values of both $\mathcal{J}_{\rm d}^{\rm LSQ}$ and the value of
\begin{equation}\label{eq:LSQfull}
    \mathcal{J}_{\rm full}^{\rm LSQ} = 
    \frac{1}{M}\norm*{\psid-e^{iS}\psi_T}^2,
\end{equation}
after a minimization over $S$, as a relative, and more familiar, measure of optimality at the level of the full-dynamical picture.

The optimal control problem~\eqref{eq:OCP} is posed over the admissible space $\mathcal{W}$, which is infinite-dimensional. We approximate this by a finite-dimensional admissible space, constructed using a Galerkin-type method called the chopped random basis method (CRAB), first used by~\cite{Caneva,calarco2000} and explained in great detail in the work of~\cite{me}. 
We use the following basis and trial functions
\begin{equation}\label{eq:CRABbasis}
    \vw_{\rm CRAB}=\vw_{\rm trial}+\vw_b\sum_{j=1}^{N_D}\frac{\epsilon_j^w}{j^2}\sin{\left(\frac{j\pi t}{T}\right)},
\end{equation}
where the value of $\vw_b$ is consistent with the boundary conditions implied by the trial controls $u_{\rm trial}$ and $v_{\rm trial}$ in Equation~\eqref{eq:trialcont}. 

The amplitudes $\varepsilon_j^w$ are random variables drawn uniformly from $[-1,1]$. We choose the coefficients $A_j=j^{-2}\epsilon_j^w$ to decay quadratically because the Fourier series of absolutely continuous functions exhibit the same type of decay~\cite{Trefethen}. In this way, the search space for optimal control ${\bf w}$ is not severely restricted, yet candidate controls remain technically feasible. To find the coefficients $\varepsilon_j^w$, we use the differential evolution (DE) method~\cite{Storn} outlined in~\ref{section:DE}. 

\subsubsection*{Remark} The numerical optimization problem associated with the objective functional in Equation~\eqref{eq:OCP} is often stated in the variational form 
of Euler-Lagrange equations and solved using a form of gradient descent~\cite{BorziBook,me}. This requires functional derivatives of the objective $\mathcal{J}$ with respect to the control vector $\vw$ and thus involves derivatives of the basis functions $\varphi_n$ in Equation~\eqref{eq:GPEGal}. Since these derivatives cannot be written in closed form, this renders gradient-based methods cumbersome. Thus we choose not to pursue such a strategy here. In previous work~\cite{me}, we numerically solve a similar optimization problem using both the CRAB method, and when possible, a combination of the CRAB method and gradient descent and find that the CRAB method alone is fairly successful in finding efficient control policies.

\section{Results of Numerical Optimization}\label{section:results}

In this section, we present the results of our numerical optimization. We briefly summarize the strategy outlined over the past sections. We perform the optimization on the three-mode reduced system described by the Hamiltonian~\eqref{eq:3mHam}. We use the initial and desired profiles for the two-mode system shown in Fig.~\ref{fig:PS1}, as we found that $c_2$ was negligibly small in the corresponding stationary solutions of the three-mode system. We minimize the phase-insensitive infidelity $\mathcal{J}$ defined in Eq.~\eqref{eq:OCP}. The control function $\vw(t)$ is constructed using the CRAB method~\eqref{eq:CRABbasis}, and the optimization is performed using the DE method described in~\ref{section:DE}. We find, through trial and error, that $N_D=15$ basis functions are sufficient for the CRAB approximation. That is, a CRAB basis~\eqref{eq:CRABbasis} with this many modes reduces the optimal control problem~\eqref{eq:OCP} to a nonlinear programming problem in $2N_D$ dimensions whose solution we find acceptable. 

\begin{figure}[htbp]
\begin{centering}
\subfigure{\includegraphics[width=0.45\textwidth]{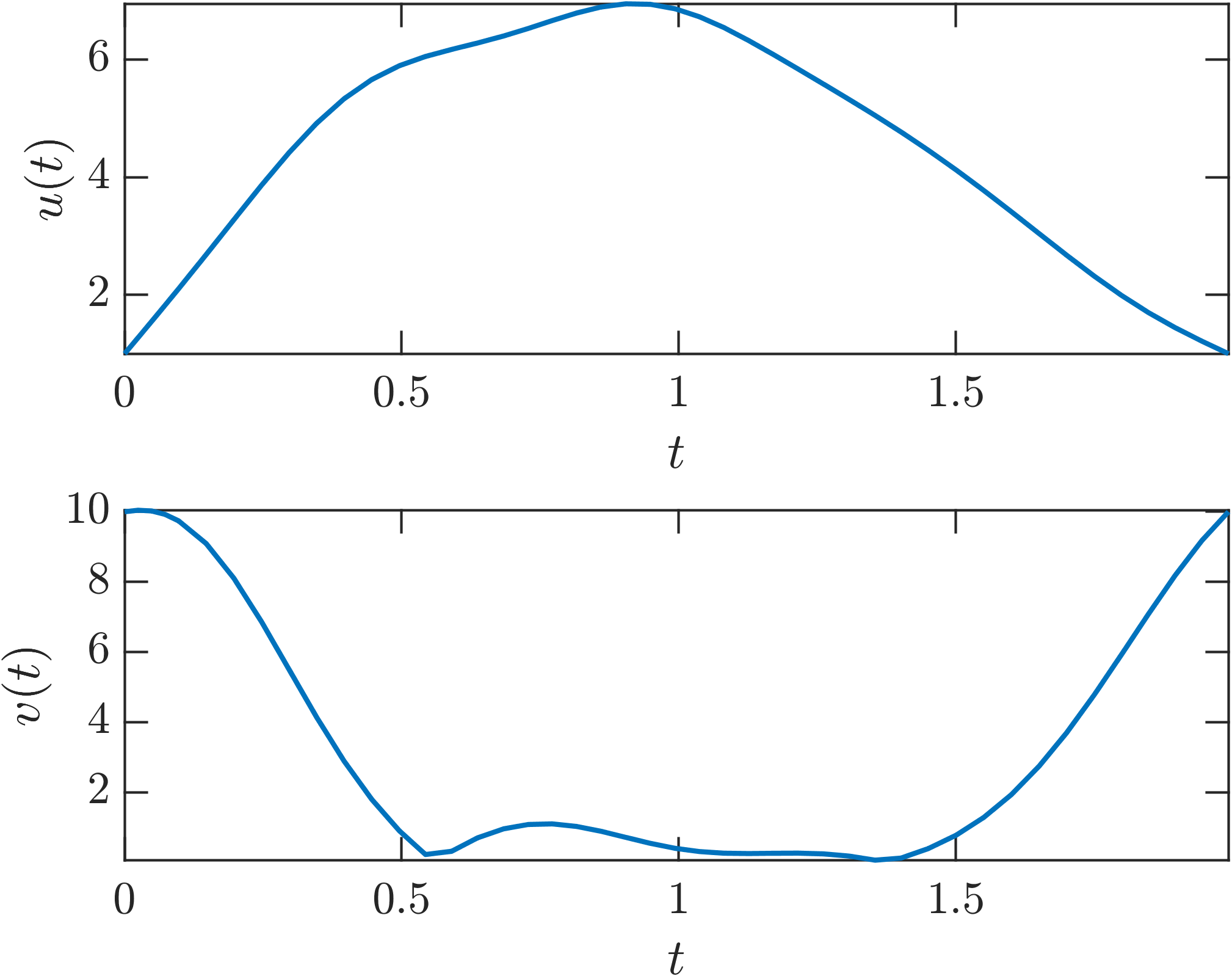}}
\subfigure{\includegraphics[width=0.45\textwidth]{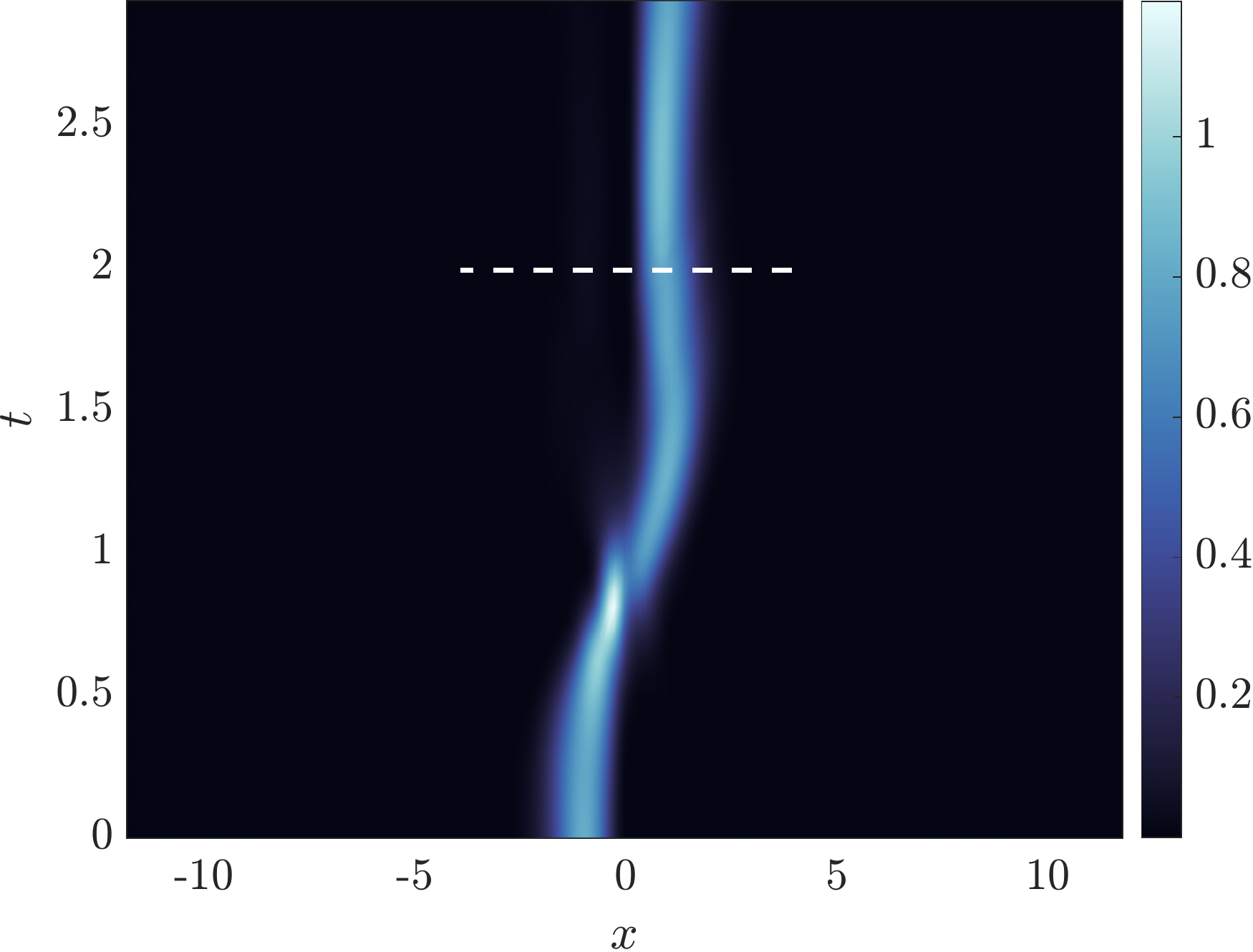}}
\subfigure{\includegraphics[width=0.45\textwidth]{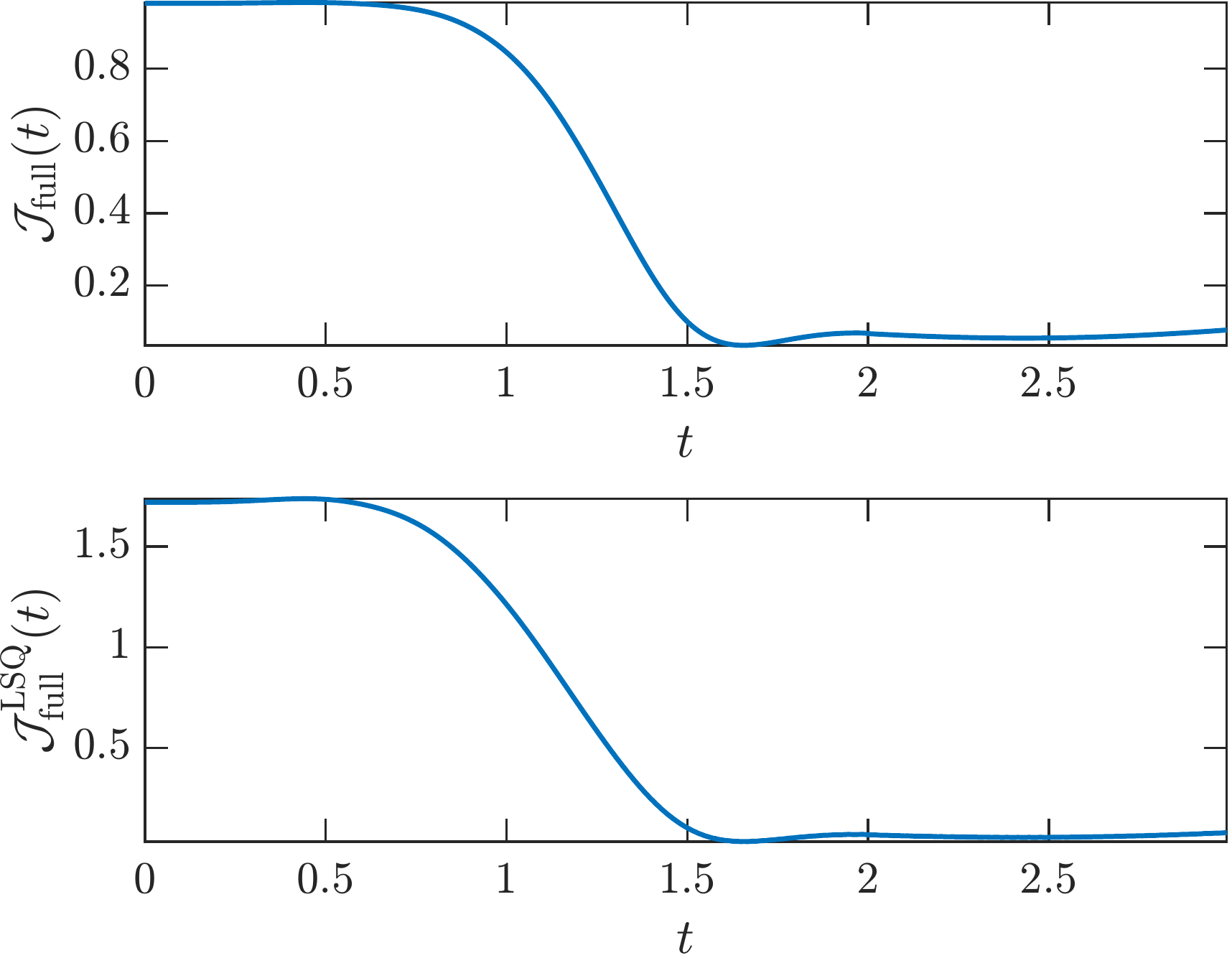}}
\subfigure{\includegraphics[width=0.45\textwidth]{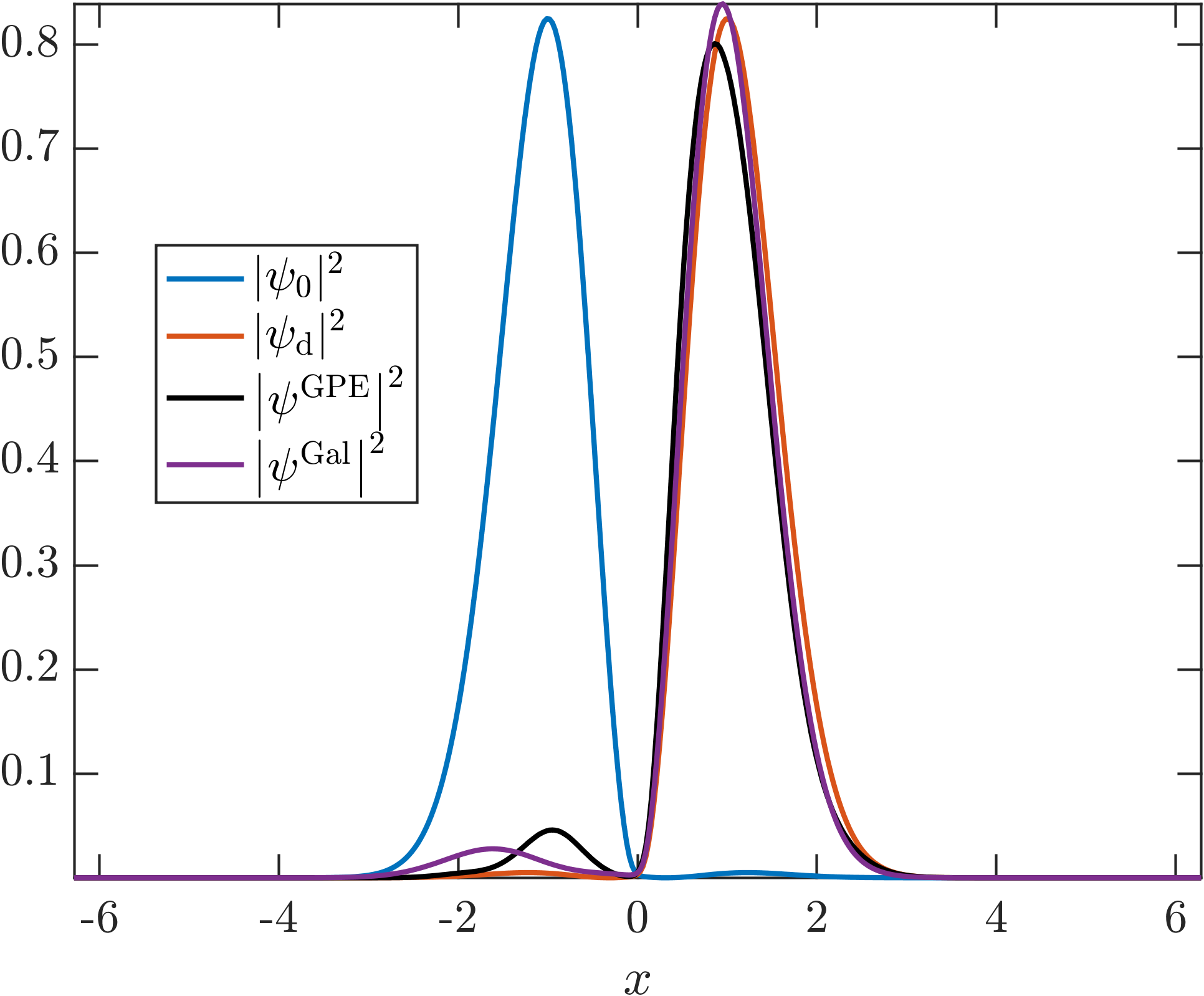}}
\caption{The result of using numerical optimal control theory. Top left: optimal controls identified via our numerical methodology. Top right: The solution of the GPE~\eqref{eq:GPE} in absolute-value squared. The dotted white line, at $T=2$, represents the moment the controls are held at their constant terminal values $\vw_b$. Bottom left: the full infidelity $\mathcal{J}_{\rm full}$ and full modified least-squares objective $\mathcal{J}^{\rm LSQ}_{\rm full}$. Bottom right: wavefunction profiles, in absolute-value squared, of the initial state $\psi_0$, the desired state $\psi_{\rm d}$ the state $\psi^{\rm Gal}$ computed via the three-mode model at $T=2$, and the state $\psi^{\rm GPE}$ computed via the GPE~\eqref{eq:GPE} at $T=2.$}\label{fig:OCresult}
\end{centering}
\end{figure}

Figure~\ref{fig:OCresult} shows the result of this numerical optimization. This computation yielded an objective function $\mathcal{J}= 0.0501$, as defined by Equation~\eqref{eq:OCP}, and a least squares infidelity $\mathcal{J}_{\rm LSQ} = 0.0422$, as defined by Equation~\eqref{eq:LSQ}.
The figure also shows the full infidelity $\mathcal{J}_{\rm full}$ and the full least-squares objective $\mathcal{J}^{\rm LSQ}_{\rm full}$ resulting from using optimal controls in simulating the GPE~\eqref{eq:GPE}.


It is important to recognize that the dynamics selected by
the optimizer somewhat conforms to the anticipated physical
intuition about the optimal strategy. It can indeed
be observed in the figure that the barrier height $v(t)$ decreases to about $0$, 
while the parabolic confinement becomes tighter, which
enables the mass to be transferred from one side of the lowered barrier to the other. 
This transfer is visible at time $t \approx 1$ in the
the top right panel of the figure. Subsequently, $v(t)$ rises sharply
again, as the parabolic confinement returns to its
original value, so that the combination of the two now ensures confinement of the 
transmitted mass to the right well. It
is through this procedure that the infidelity is substantially
decreased in the lower left panel, and indeed subsequently remains
small, during the return of the confinement conditions to their
original settings.


\begin{figure}[htbp]
\begin{centering}
\subfigure{\includegraphics[width=0.45\textwidth]{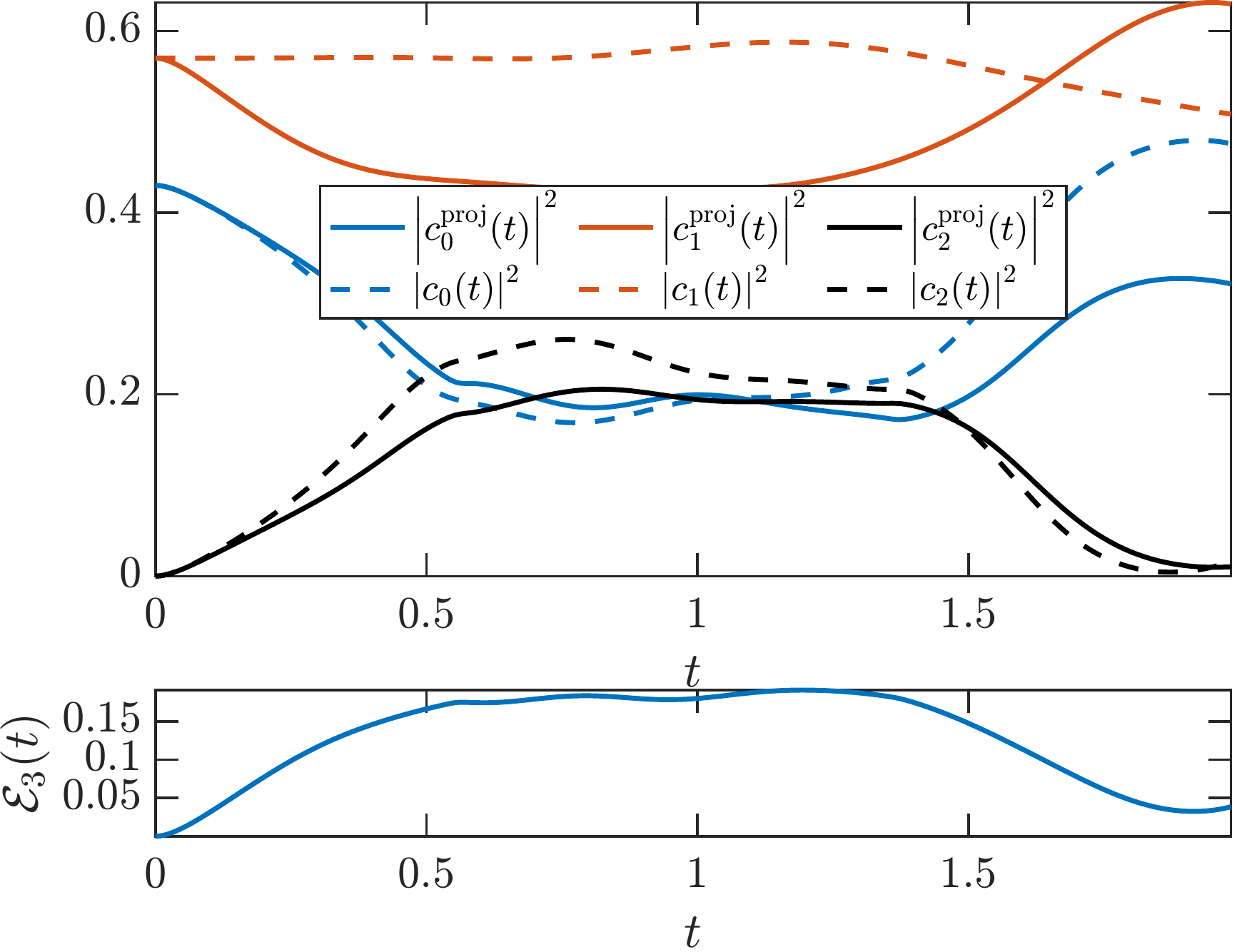}}
\subfigure{\includegraphics[width=0.45\textwidth]{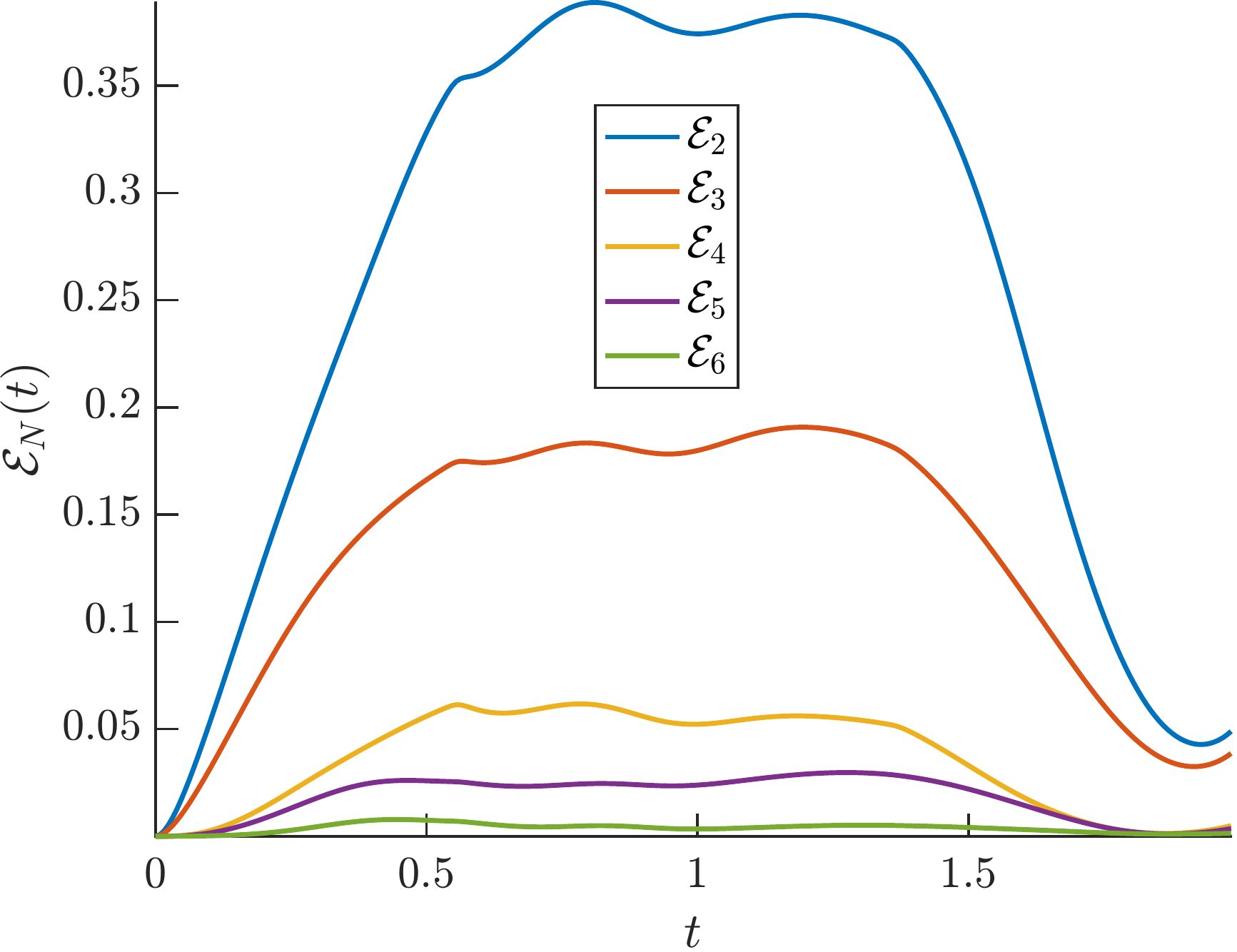}}
\caption{Left: Numerical solution of the three-mode model, projected coefficients, and the error consistent with Figure~\ref{fig:OCresult}. Right: the error of modes two through six, with the natural number $N$ used consistently with the Galerkin reduction defined by Equation~\eqref{eq:GPEGal}.}\label{fig:Energyanalysis}
\end{centering}
\end{figure}

We also show, in Figure~\ref{fig:Energyanalysis}, an error analysis. The left panel in Figure~\ref{fig:Energyanalysis} is equivalent to the lower-left panel of Figure~\ref{fig:3mT}, showing the coefficients $c_j(t)$ and the relative error $\mathcal{E}_3(t)$ defined by Equation~\eqref{eq:relerr}. While the $\max_t \mathcal{E}_3(t)$ is lower in the simulation using the optimal control, than in the trial control, it is interesting to note that using the optimal control yields significantly better agreement between the dynamics of the full GPE system and its projection onto the first three modes at the final time $T$, decreasing from a maximum of about 18\% to just about 5\% at time $T$. The right panel of the figure computes the error $\mathcal{E}_N(t)$ for values of $N\le6$. We find $\mathcal{E}_N(t)$ decreases monotonically, pointwise in $t,$ as expected. Although we have shown that three modes suffice to control the GPE, this result indicates the expected accuracy for $N>3$. 




\section{Conclusions}\label{section:conclusions}

In the present work, we have applied the methodology of optimal
control to dynamics in double-well potentials,  one of the most prototypical (and highly controlled) experimental settings,
both in atomic Bose-Einstein condensates and in nonlinear optics.
We have adapted the relevant methodology to the well-established description
of Galerkin truncations within this setting. Typically, two-mode truncations
are used in order to explore the steady states, stability, bifurcations, and
dynamics of such double-well systems.  A key finding of the present work
is that such a two-mode approximation is not sufficient in order to 
characterize the dynamics associated with optimal transport. 

Indeed, it was found that the involvement of a three-mode approximation was crucial in order to adequately describe the relevant dynamical process,
identified as optimal by the differential evolution algorithm deployed herein.
Despite  quantitative inaccuracies in using a low-order Galerkin reduction, we see that controlling the three-mode model effectively controls the GPE wavefunction in numerical simulations. Furthermore, our numerical results showcase the extent to which the model can be refined by taking the Galerkin reduction out to higher orders 
(recall that the basis we use is complete in $L^2$, 
hence the Galerkin representation converges to the wavefunction $\psi$ satisfying the GPE in $L^2$ as the number of modes N increases). In particular, our computations, shown in Figure~\ref{fig:Energyanalysis}, indicate that substantial gain can be made by including a fourth mode into the model. We leave the potential inclusion of these higher modes in a reduction-based optimal control strategy as a subject for future work.

Additionally, the adaptation of such optimal control methodologies to 
low-dimensional truncated Galerkin dynamics is a technique that
could find significant potential for further
applications. Some possibilities include multi-component and
spinor condensates~\cite{KAWAGUCHI2012253} where few-mode approximations
have proven useful~\cite{WANG20082922,gunay,tian}.
Moreover, extending such control strategies beyond the mean-field
framework and into the realm of many-body effects~\cite{masiello,polls},
is of particular interest in its own right; for a review of the latter,
see, e.g., the recent preprint of~\cite{mistakidis2022cold}. In regards to complementary or alternative low-dimensional control strategies, it should be possible to couple the optimal control methods used here with methods based on the so-called shortcut to adiabaticity ~\cite{RevModPhys.91.045001}. 
We additionally note that other 
methods more tailored to the driving 
of specific potentials~\cite{bromley}, as
well as ones that are model 
agnostic~\cite{ML} also exist in the literature and are of interest in their own right.

It is also natural to extend these ideas to higher-dimensional systems where few-well arrangements have also been explored~\cite{PhysRevE.80.046611} or to systems with disorder, such as recent work which makes use of deep learning~\cite{PhysRevApplied.17.024040}. The Galerkin approach here may be generalized to systems with disorder so that optimal control policies can be computed via a low-dimensional dynamic program. Such studies are currently under consideration and will be presented in future publications.

\section*{Acknowledgments}
This material is based upon work supported by the US
National Science Foundation under Grants No. DMS-1809074 and
PHY-2110030 (P.G.K.).

\section*{Disclosures} The authors declare no conflicts of interest.

\section*{Data Availability} Data underlying the results presented in this paper are not publicly available at this time but may be obtained from the authors upon reasonable request.

\bibliographystyle{siam} 
\bibliography{bibliography.bib} 

\appendix
\section{Numerical Method for Solving the Gross-Pitaevskii Equation}\label{section:GPENum}
It is necessary to solve the GPE~\eqref{eq:GPE} in order to validate the Galerkin reduction~\eqref{eq:GPEGal} and to evaluate the performance of the dimensionally reduced optimal control problem. The boundary conditions are assumed to be periodic so that the use of spectral methods is straightforward. We use an operator splitting method, and, to this end, rewrite Equation~\eqref{eq:GPE} in the form
\begin{equation}\label{eq:splitpsi}
i\partial_t\psi=\mathcal{L}\psi+\mathcal{N}(\psi),
\end{equation}
where the linear operator is given by $\mathcal{L}=-\frac{1}{2}\partial_x^2$ and the nonlinear and inhomogeneous operator $\mathcal{N}(\psi)$ incorporates the remaining terms. 

We choose to use a second order in time operator splitting, often referred to as Strang splitting~\cite{Strang}, to approximate the resulting matrix exponential by
\begin{equation}\label{eq:strangsplit}
e^{\left(\mathcal{L}+\mathcal{N}\right)Mh}=e^{\mathcal{L}h/2}e^{\mathcal{N}h}e^{\mathcal{L}h}\ldots e^{\mathcal{L}h}e^{\mathcal{N}h}e^{\mathcal{L}h/2}+\mathcal{O}\left(h^2\right),
\end{equation}
where $h=T/M$ is the time discretization for a given number of time steps $M$. The solution of the linear equation resulting from the matrix exponential $e^{\mathcal{L}h}$ is facilitated by the Fourier transform and is given by
\begin{equation}
\psi_{n+1}=\mathcal{F}^{-1}\left\{\mathcal{F}\{\psi_n\}e^{-\frac{ihk^2}{2}}\right\},
\end{equation}
where $\mathcal{F}$ and $\mathcal{F}^{-1}$ denote the Fourier and inverse Fourier transformation, respectively. Fourier and inverse Fourier transforms are computed via the fast Fourier transform functions in MATLAB, \texttt{fftw} and \texttt{ifftw}, with discretized wavenumbers $k\in\left[\frac{-N/2+1}{2l},\frac{N/2-1}{2 l}\right],$ where $2\pi l$ units of length are assumed in the truncated spatial domain and $N$ is the number of spatial discretization points. 

The nonlinear equation resulting from the matrix exponential $e^{\mathcal{N}h}$ is quite simple. Since the nonlinearity does not involve spatial derivatives, we are simply tasked with solving ODEs. Using polar coordinates, i.e., letting
\begin{equation}
\psi=\rho(x,t)e^{i\theta(x,t)},
\end{equation}
results in the system
\begin{equation}\label{eq:gpenumtheta}
\begin{alignedat}{2}
    \partial_t\rho&=0,\\
    \partial_t{\theta}&=-\rho^{2}-V(x,t).
\end{alignedat}
\end{equation}
The first of these equations is a statement about conservation of the mass $||\psi||_{L^2({\mathbb{C}})}$. The second equation, governing the phase $\theta,$ can be solved via
any number of standard numerical ODE techniques; we simply use the second-order accurate midpoint method. The update for the phase $\theta,$ in this case, is
\begin{equation}
    \theta(x,t_{n+1})=\theta(x,t_{n})-h\rho(x,t_n)^{2}-hV\left(x,t_n+\frac{h}{2}\right)
\end{equation}

\section{Details of the Three-Mode Model}\label{section:3m}

The Hamiltonian for the three-mode model is found to be
\begin{align}\label{eq:3mHam}
    \mathcal{H}&=\alpha\abs{c_0}^2+\beta\abs{c_1}^2+\sigma\abs{c_2}^2+2\Delta\Im\left\{c_0\bar{c}_2\right\}+\frac{\gamma_0}{2}\abs{c_0}^4+\frac{\gamma_1}{2}\abs{c_1}^4+\frac{\gamma_2}{2}\abs{c_2}^4  \\
    &+\gamma_3\left(\Re\left\{c_0^2 \bar{c}_1^2\right\}+2\abs{c_0}^2\abs{c_1}^2\right)+\gamma_4\left(\Re\left\{c_0^2 \bar{c}_2^2\right\}+2\abs{c_0}^2\abs{c_2}^2\right)+\gamma_5\left(\Re\left\{c_1^2 \bar{c}_2^2\right\}+2\abs{c_1}^2\abs{c_2}^2\right)\nonumber\\
    &+2\left(\gamma_6\abs{c_0}^2+\gamma_7\abs{c_2}^2\right)\Re\left\{c_0\bar{c}_2\right\}+2\gamma_8\left(\Re\left\{c_0\bar{c_1}^2c_2\right\}+2\Re\left\{c_0\abs{c_1}^2\bar{c}_2\right\}\right),
\end{align}
where the projection coefficients are given by
\begin{align}
&\alpha=\ip{\mathcal{L}\varphi_0}{\varphi_0},\  
\beta=\langle{\mathcal{L}\varphi_1,\varphi_1}\rangle,\ 
\sigma=\ip{\mathcal{L}\varphi_2}{\varphi_2},\ 
\Delta=\ip{\varphi_0}{\dot{\varphi}_2}, \nonumber \\
&\gamma_0=\|\varphi_0^4\|,\ \gamma_1=\|\varphi_1^4\|,\ 
\gamma_2=\|\varphi_2^4\|,
\gamma_3=\ip{\varphi_0^2}{\varphi_1^2},\ 
\gamma_4=\ip{\varphi_0^2}{\varphi_2^2}, \nonumber \\
&\gamma_5=\ip{\varphi_1^2}{\varphi_2^2},\
 \gamma_6=\ip{\varphi_0^3}{\varphi_2},\
\gamma_7=\ip{\varphi_0}{\varphi_2^3},
\gamma_8=\ip{\varphi_0}{\varphi_1^2\varphi_2}.
\end{align}
from which the analogous system to the two-mode system~\eqref{eq:2msys} can be derived easily. The appearance of the term with coefficient $\Delta$ is due to the time-dependent nature of the basis functions $\varphi_n$, and did not appear in the two-mode system due to the parity of the first two modes. This, as well as the implicit claim that $\ip{\varphi_0}{\dot{\varphi}_2}=-\ip{\varphi_2}{\dot{\varphi}_0}$ may be verified numerically.
 
As was done with the two-mode system, we use changes of variables and the, now three-mode, discrete mass to reduce dimensionality. Because we may only reduce the number of degrees of freedom to two for the three-mode system, we lose the comfort of visualizing the dynamics via phase portraits as before. Nevertheless, we pursue a reduction since this helps make identifying fixed points of the dynamical system simpler. To this end, we begin by using

\begin{equation}\label{eq:fourthcan}
c_0=\eta_0e^{i\theta},\quad 
c_1=re^{i\theta},\quad
c_2=\eta_1e^{i\theta},
\end{equation}
where $\eta_j\in\CC$ and $r\in\RR.$ We can easily eliminate $r$ by using the discrete mass once again: $r=\sqrt{\Md-\abs{\eta_0}^2-\abs{\eta_1}^2}.$ Now, converting to polar coordinates via
\begin{equation}\label{eq:fifthcan}
    \eta_0= \sqrt{\rho_0}e^{i \theta_0},\quad\eta_1= \sqrt{\rho_1}e^{i\theta_1} ,
\end{equation}
we find the following, two-degree of freedom Hamiltonian

\begin{align}\label{eq:3mPolHam}
    \mathcal{H}=&\frac{\Md }{2}\left(2 \beta +\gamma_1 \Md\right)+\frac{\rho_0^2 }{2}\left(\gamma_0+\gamma_1-2 \gamma_5 \left(\cos \left(2 \theta_0\right)-2\right)\right)-\frac{\rho_1^2}{2}\left(\gamma_1+\gamma_2-2 \gamma_7 \left(\cos \left(2 \theta_1\right)-2\right)\right)\nonumber \\
    &\rho_0 \left(\alpha -\beta -\gamma_1 \Md+\gamma_5 \Md\left(2+ \cos \left(2 \theta_0\right)\right) \right)+\rho_1 \left(\beta+\gamma_1 \Md-\gamma_7 \Md\left(2+ \cos \left(2 \theta_1\right)\right)-\sigma \right)\nonumber \\
    &+\rho_0 \rho_1 \left(\gamma_1-\gamma_5 \left(2+\cos \left(2 \theta_0\right)\right)+\gamma_6\left(2+ \cos \left(2 \left(\theta_0-\theta_1\right)\right)\right)-\gamma_7 \left(2+\cos \left(2 \theta_1\right)\right)\right)\nonumber \\
    &+2 \rho_0^{3/2} \rho_1^{1/2} \left(\gamma_3 \sin \left(\theta_0\right) \sin \left(\theta_1\right)-\gamma_8 \sin \left(\theta_0\right) \sin \left(\theta_1\right)+\gamma_3 \cos \left(\theta_0\right) \cos \left(\theta_1\right)-3 \gamma_8 \cos \left(\theta_0\right) \cos \left(\theta_1\right)\right) \nonumber \\
    &+2\rho_0^{1/2} \rho_1^{3/2} \left(\gamma_4 \sin \left(\theta_0\right) \sin \left(\theta_1\right)-\gamma_8 \sin \left(\theta_0\right) \sin \left(\theta_1\right)+\gamma_4 \cos \left(\theta_0\right) \cos \left(\theta_1\right)-3 \gamma_8 \cos \left(\theta_0\right) \cos \left(\theta_1\right)\right)\nonumber \\
    &+2 \Md \gamma_8 \sqrt{\rho_0\rho_1}  \left(\sin \left(\theta_0\right) \sin \left(\theta_1\right)+3 \cos \left(\theta_0\right) \cos \left(\theta_1\right)\right).
\end{align}

\section{Optimization via Differential Evolution}\label{section:DE}
DE is a stochastic optimization method used to search for candidate solutions to non-convex optimization problems. The idea behind DE is a so-called genetic algorithm that draws inspiration from evolutionary genetics. DE searches the space of candidate solutions by initializing a population set of vectors, known as agents, within some chosen region of the search space. These vectors are then randomly mutated into a new population set, or generation. 

 \begin{algorithm}[htbp]
\caption{Differential Evolution Mutation}\label{algo:mut}
\KwResult{A vector $z$ mutated from agents in a given generation as required by the DE Algorithm~\eqref{algo:HDE}.}
\SetKwInOut{Input}{Input}
 \Input{4 distinct members $a,b,c,d$ from the current generation of agents each with $N$ components, the crossover ratio $R_C\in(0,1)$, and weight $F\in(0,2)$.}
 \For{j=1:N}{
              Compute a random variable $\mathtt{rand}$\;
                 \eIf {$\mathtt{rand}<R_C$} {
                     $z[j]\gets a[j]+F*(b[j]-c[j])$
                     }
                     {$z[j]\gets d[j]$
                 }
         }
\end{algorithm}

 \begin{algorithm}[htbp]
\caption{Differential Evolution}\label{algo:HDE}
\KwResult{A vector likely to be globally optimal with respect to an objective $J$.}
\SetKwInOut{Input}{Input}
 \Input{A maximum number of iterations $\mathtt{Nmax}$, crossover ratio $R_C\in(0,1)$ and weight $F\in(0,2)$}
\While{$\mathtt{counter}<\mathtt{Nmax}$}{
Generate a population \texttt{pop} of $N_{\rm pop}$ vectors.

 \For{$i=1:N_{\rm pop}$}{
 $\mathtt{CurrentMember}\gets \mathtt{Pop}_i$\;
     Choose three distinct vectors $a_i,b_i,c_i$ different from the vector $\mathtt{Pop}_i$\;
        Mutate $a_i,b_i,c_i$, and the $\mathtt{CurrentMember}$ into the mutated vector $z$
        using the mutation parameters $R_C,F$ and Algorithm~\ref{algo:mut}\;
         \If{$J(z)<J(\mathtt{CurrentMember})$} {
             $\mathtt{TemporaryPop}_i=z$\;
         }
 }
 $\mathtt{Pop}\gets \mathtt{TemporaryPop}$\;
 $\mathtt{counter}\gets \mathtt{counter}+1$\;
}
 \end{algorithm}
 
The mutation operates via two mechanisms: a weighted combination and a "crossover" which randomly exchanges "traits", or vector elements, between agents. The method requires three parameters; the weight $F\in(0,2)$, the crossover parameter $R_C\in(0,1)$, and the size of the population $N_{\rm pop}$, which, by Algorithm~\eqref{algo:mut}, is required to be an integer greater than three. 
A pseudo-code illustrating the implementation of
the relevant algorithms is given in Algorithms~\eqref{algo:mut}
and~~\eqref{algo:HDE}.
Through trial and error, we find the parameters $F=0.8,\ R_C=0.9,\ N_{\rm pop}=20$ work well.
DE ensures that the objective functional $\mathcal{J}$ decreases monotonically with each generation. As each iteration "evolves" into the next, inferior vectors "inherit" optimal traits from superior vectors via mutations. DE only allows mutations that are more optimal with respect to $\mathcal{J}$ to pass to the next generation. After a sufficient number of iterations, the best vector in the final generation is chosen as the candidate solution most likely to be globally optimal with respect to an objective functional.

Genetic algorithms, which require very few assumptions about the objective functional, are part of a wider class of optimization methods called metaheuristics. Although metaheuristics are useful for non-convex optimization problems, they do not provide
guarantees about the global optimality of candidate solutions. Since the algorithm is stopped after a finite number of iterations, different random realizations return different candidate optimizers. The results we show are the best among five different realizations. 

In practice, we do not recommend taking fewer realizations since one runs the risk of computing highly sub-optimal controls, which, indeed, is a generic issue when solving non-convex optimization problems using stochastic methods. Of course, taking more realization is expensive, but we find that with the optimization and physical parameters used throughout this work, five realizations are sufficient to guarantee the discovery of, at least, a couple of  optimal control policies which are extremely competitive with regards to the objective functional~\eqref{eq:OCP}. From a computation of 30 realizations, about a third of the control policies are within 1\% of the infidelity returned by the best control.

\end{document}